\documentclass[aps,prx,showpacs,twocolumn,amsmath,amssymb,superscriptaddress,nofootinbib]{revtex4}
\usepackage{graphicx}
\usepackage{bm}
\usepackage{amssymb}
\usepackage{amsmath}
\usepackage{latexsym}
\usepackage{color}
\newcommand{\vc}[1]{\mbox{\boldmath $#1$}} 
\newcommand{\ind}[1]{_{#1}}    
\newcommand{\indrm}[1]{_{\mathrm {#1}}}    
\newcommand{\dei}[1]{\Delta E_{\ind{#1}}}   
\newcommand{\dai}[1]{\Delta \theta_{\ind{#1}}}   
\newcommand{\dirate}{{\mathcal D}}   
\newcommand{\sgn}{s}   
\newcommand{\dcomm}[1]{b_{\cup_{\ind{#1}}}}
\newcommand{\acomm}[1]{1/b_{\cup_{\ind{#1}}}}
\newcommand{\fcomm}[1]{\dirate_{\cup_{\ind{#1}}}}

\newcommand{\crystal}{C}

\newcommand{\uhrix}{U}
\newcommand{\crl}{lens}
\newcommand{\crls}{CRL}
\newcommand{\focus}{F}
\definecolor{bluish}{rgb}{0.3,0.3,0.7}
\begin{document}

  \title{Novel opportunities for sub-meV inelastic X-ray scattering at
    high-repetition rate self-seeded X-ray free-electron lasers }

\author{Oleg Chubar}
\affiliation{National Synchrotron Light Source II, Brookhaven National Laboratory, Upton NY 11973, USA}
\author{Gianluca Geloni}
\affiliation{European X-ray Free-Electron Laser, 22761 Hamburg, Germany}
\author{Vitali Kocharyan}
\affiliation{Deutsches Elektronen-Synchrotron, 22607 Hamburg, Germany}
\author{Anders Madsen}
\affiliation{European X-ray Free-Electron Laser, 22761 Hamburg, Germany}
\author{Evgeni Saldin}
\affiliation{Deutsches Elektronen-Synchrotron, 22607 Hamburg, Germany}
\author{Svitozar Serkez}
\affiliation{Deutsches Elektronen-Synchrotron, 22607 Hamburg, Germany}
\author{Yuri Shvyd'ko}
\email{shvydko@aps.anl.gov}
\thanks{corresponding author}
\affiliation{Advanced Photon Source, Argonne National Laboratory, Argonne Il 60439, USA}
\author{John Sutter}
\affiliation{Diamond Light Source Ltd, Didcot OX11 0DE, United Kingdom}

\begin{abstract}
  Inelastic X-ray scattering (IXS) is an important tool for studies of
  equilibrium dynamics in condensed matter. A new spectrometer
  recently proposed for ultra-high-resolution IXS (UHRIX) has achieved
  0.6~meV and 0.25~nm$^{-1}$ spectral and momentum transfer
  resolutions, respectively. However, further improvements down to
  0.1~meV and 0.02~nm$^{-1}$ are required to close the gap in
  energy-momentum space between high and low frequency probes. We show
  that this goal can be achieved by further optimizing the X-ray
  optics and by increasing the spectral flux of the incident X-ray
  pulses. UHRIX performs best at energies from 5 to 10 keV, where a
  combination of self-seeding and undulator tapering at the SASE-2
  beamline of the European XFEL promises up to a hundred-fold increase
  in average spectral flux compared with nominal SASE pulses at
  saturation, or three orders of magnitude more than what is possible
  with storage-ring based radiation sources. Wave-optics calculations
  show that about $7\times 10^{12}$~ph/s in a $90$-$\mu$eV bandwidth
  can be achieved on the sample. This will provide unique new
  possibilities for dynamics studies by IXS.
\end{abstract}


\maketitle

\section{\label{sec:intro} Introduction}

\begin{figure}[t!]
\setlength{\unitlength}{\textwidth}
\begin{picture}(1,0.49)(0,0)
\put(0.0,0.00){\includegraphics[width=0.50\textwidth]{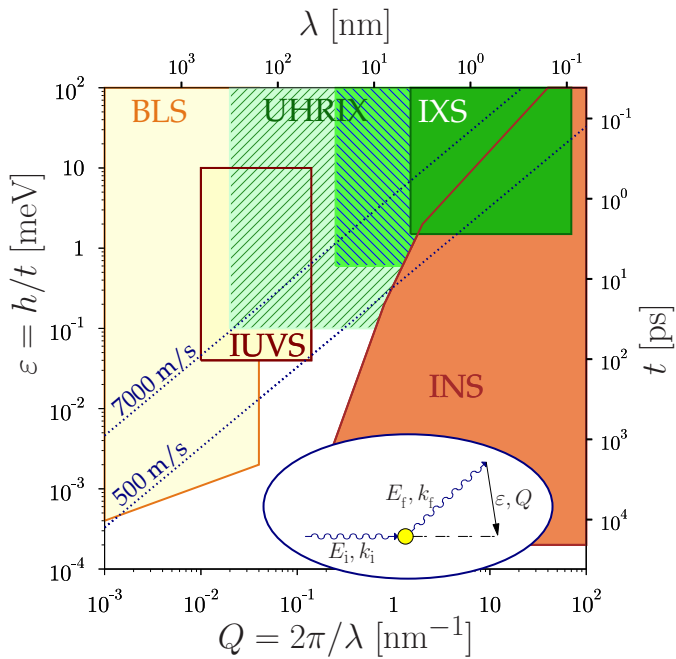}}
\end{picture}
\caption{Time-length ($t-\lambda$) and energy-momentum
  ($\varepsilon\!-\!Q$) space of excitations relevant in condensed
  matter. The figure indicates how different domains are accessed by
  different inelastic scattering probes: neutrons (INS), X-rays (IXS),
  ultraviolet (IUVS), and Brillouin Light Scattering (BLS). The
  ultra-high-resolution IXS (UHRIX) spectrometer presented in
  Ref.~\cite{SSS14} entered the previously inaccessible region marked
  in shaded green. The novel capabilities discussed in the present
  paper will enable IXS experiments with 0.1-meV and 0.02-nm$^{-1}$
  resolution in the region marked in shaded light green. Hence, it
  allows to close the existing gap between the high-frequency and
  low-frequency probes. The energy
  $\varepsilon=E_{\indrm{f}}\!-\!E_{\indrm{i}}$ and the momentum
  $\vc{Q}=\vc{k}_{\indrm{f}}\!-\!\vc{k}_{\indrm{i}}$ transfers from
  initial to final photon/neutron states are measured in inelastic
  scattering experiments, as schematically shown in the inset.}
\label{fig000}
\end{figure}

Momentum resolved inelastic X-ray scattering (IXS) is a technique
introduced \cite{BDP87,Burkel} and widely used
\cite{Sette98,Burkel2,KS07,MonacoIXS15,Baron15arxiv} at synchrotron
radiation facilities for studies of atomic-scale dynamics in condensed
matter. IXS is a photon-in-photon-out method applicable to any
condensed matter system, whether it is solid, liquid, biological, or
of any other nature. A photon with energy $E_{\indrm{i}}$ and momentum
$\vc{K}_{\indrm{i}}$ changes its energy and momentum to
$E_{\indrm{f}}$ and $\vc{K}_{\indrm{f}}$ in an inelastic scattering
process in the sample and leaves behind a collective excitation with
energy $\varepsilon=E_{\indrm{i}}-E_{\indrm{f}}$ and momentum
$\vc{Q}=\vc{K}_{\indrm{i}}-\vc{K}_{\indrm{f}}$, respectively, as shown
in the sketch in Fig.~\ref{fig000}. The interpretation of IXS is
straightforward as it measures the dynamical structure factor
$S(\vc{Q},\varepsilon)$, {\em i.e.} the spatiotemporal Fourier
transform of the van Hove time-dependent pair correlation function
\cite{Askcroft}. Therefore, it provides access to dynamics on a length
scale $\lambda=2\pi/Q$ and at a time scale $t=2\pi\hbar/\varepsilon$.

IXS is one of only a few existing inelastic scattering
techniques. Each technique provides access to a limited region in the
time-length scale or equivalently in the energy-momentum space of
collective excitations relevant for condensed
matter. Figure~\ref{fig000} shows how a broad range of excitations are
covered by different inelastic scattering probes: neutrons (INS),
X-rays (IXS), ultraviolet light (IUVS), and Brillouin Light Scattering
(BLS). A gap remains in experimental capabilities between
low-frequency (visible and ultraviolet light) and high-frequency
(X-rays and neutrons) inelastic scattering techniques. Hence, dynamics
in the range from about 1-100 picosecond (ps) on atomic- and
meso-scales is still inaccessible by any known experimental
probe. This is precisely the region of vital importance for disordered
systems and therefore many outstanding problems in condensed matter
dynamics, such as the nature of the liquid to glass transition, could
be addressed by entering this unexplored domain.

In principle there are no limitations preventing IXS from penetrating
this unexplored dynamic range of excitations\footnote{INS cannot enter
  this region due to the kinematic limitation. The low-frequency
  probes cannot enter this region because their photon wavelengths are
  too long.}.  This would, however, require solving two longstanding
challenges in IXS. First, IXS spectrometers in their traditional
implementation rely on an X-ray optics concept utilizing single-bounce
Bragg back-reflecting spherical analyzers, leading to pronounced
Lorentzian tails of the spectral resolution function. This approach
has reached an impasse where the best numbers in energy ($\simeq
1.5$~meV) and momentum transfer ($\simeq 1.5$~nm$^{-1}$) resolutions
have not improved for the past 20 years \cite{MBKRSV96,SSD11}. Second,
the IXS signal is very weak. For example, with $\simeq 10^9$ incident
photons there is often less than one photon inelastically scattered
into the detector. Hence, more efficient IXS spectrometers with better
resolution and more powerful X-ray sources are required to advance the
field.

Recently, a new type of dispersive spectrometer was tested for the
first time.  This ultra-high-resolution IXS (UHRIX) spectrometer
\cite{SSS14} achieved a spectral resolution of 0.6~meV at a momentum
transfer down to 0.25~nm$^{-1}$ (shaded green area in
Fig.~\ref{fig000}). Additionally, the spectral contrast improved by an
order of magnitude compared to traditional IXS spectrometers
\cite{BDP87,SRK95,MBKRSV96,Baron1,SAB01,SSD11}. To sharpen the desired
resolution to 0.1~meV and 0.02-nm$^{-1}$ and to ensure higher count
rates, we propose to further develop the angular dispersive X-ray
optical scheme \cite{SSM13,SSS13} replacing scanning IXS spectrometers
with broadband imaging spectrographs \cite{Shvydko15} \footnote{A
  Fourier-transform IXS technique has been demonstrated recently
  \cite{TFC13}, which can be considered as a powerful complementary
  approach for studies of {\em non-equilibrium} excitations with
  ultra-high spectral resolution.}.

In addition to these optics developments, new types of X-ray sources
are on the horizon that will overcome the problem of insufficient IXS
cross-section by delivering a higher spectral flux, namely seeded
high-repetition rate X-ray free-electron lasers (XFELs). Low-gain
X-ray free-electron laser oscillators (XFELOs) may in some time in the
future produce a spectral flux of up to $10^{14}-10^{15}$
photons/s/meV \cite{KSR08,LSKF11}, but currently they are still under
conceptual development \cite{XFELO-LCLSII}. High-gain XFELs, on the
other hand, are available today. Self-amplified spontaneous emission
(SASE) X-ray free-electron lasers (XFELs) \cite{EAA10,SACLA,EXFEL-TDR}
deliver light pulses with unprecedented peak power compared to
storage-ring based sources. However, the average photon flux that can
be delivered is limited due to the low repetition rate of their linac
drivers. By contrast, the European XFEL's plan to adopt
superconducting accelerator technology will allow for producing
$27000$ X-ray pulses per second, {\em i.e.} orders of magnitude above
the $120$ pulses per second of the LCLS and the $60$ pulses per second
at SACLA.

The UHRIX instrument with the desired 0.1~meV resolution can be
installed at the SASE-2 beamline of the European XFEL together with
the MID instrument \cite{MID13} operating in the 5-25 keV range. UHRIX
performs best at relatively low photon energies between $5$ and $10$
keV with an optimum around $9$ keV. Owing to the high repetition rate
of the European XFEL, the nominal average output flux at SASE-2
amounts to about $10^{12}$ photons/s/meV at 9 keV, which is more than
one order of magnitude greater than at synchrotron radiation
facilities \cite{Baron15arxiv}. Furthermore, the spectral flux can be
substantially increased by self-seeding \cite{GKS11,HXRSS12}, which at
the European XFEL first will be available at the SASE-2 beamline
\cite{SXSEED}. Another order of magnitude increase in flux is
achievable by tapering the magnetic field of the seeded undulator
\cite{STC79,KMR81,OAC86,Fawley2002537,WFH09,GiKoSa10,FFH11,JWC12}. We
therefore propose an optimized configuration of the SASE-2 X-ray
source combining self-seeding and undulator tapering techniques in
order to reach more than $10^{14}$ photons/s/meV, the same number
estimated in Ref.~\cite{YS13}. In combination with the advanced IXS
spectrometer described here, this may become a real game-changer for
ultra-high-resolution X-ray spectroscopy, for IXS in particular, and
hence for the studies of dynamics in disordered systems.

The paper is organized as follows: in Section~\ref{sec:SASE-2} we
demonstrate that self-seeding, combined with undulator tapering,
allows achieving the aforementioned figure of $10^{14}$ photons per
second per meV bandwidth at the optimal photon energy range around 9
keV. This result is achieved by careful numerical modeling using the
XFEL code GENESIS \cite{GENESIS} and start-to-end simulations for the
European XFEL. In Section~\ref{xrayoptics} we introduce and evaluate
the X-ray optical design to achieve 0.1~meV resolution IXS. The choice
of optical elements and their design parameters are studied by
dynamical theory calculations for monochromatization in\
Section~\ref{monos}, and by geometrical optics considerations for
X-ray focusing in Section~\ref{ray}. The spectrograph design with a
spectral resolution of 0.1~meV in a 5.8~meV wide spectral window of
imaging is presented in Section~\ref{spectrograph}. The design
parameters are verified in Section~\ref{wave} by wavefront propagation
simulations from source to sample using a combination of GENESIS
\cite{GENESIS} and SRW \cite{SRW} codes. All results are summarized
and discussed in Section~\ref{conclusions}.

\section{\label{sec:SASE-2} High average flux X-ray source for Ultra-High-Resolution IXS}
\subsection{Concept}

\begin{figure}
\includegraphics[width=0.5\textwidth]{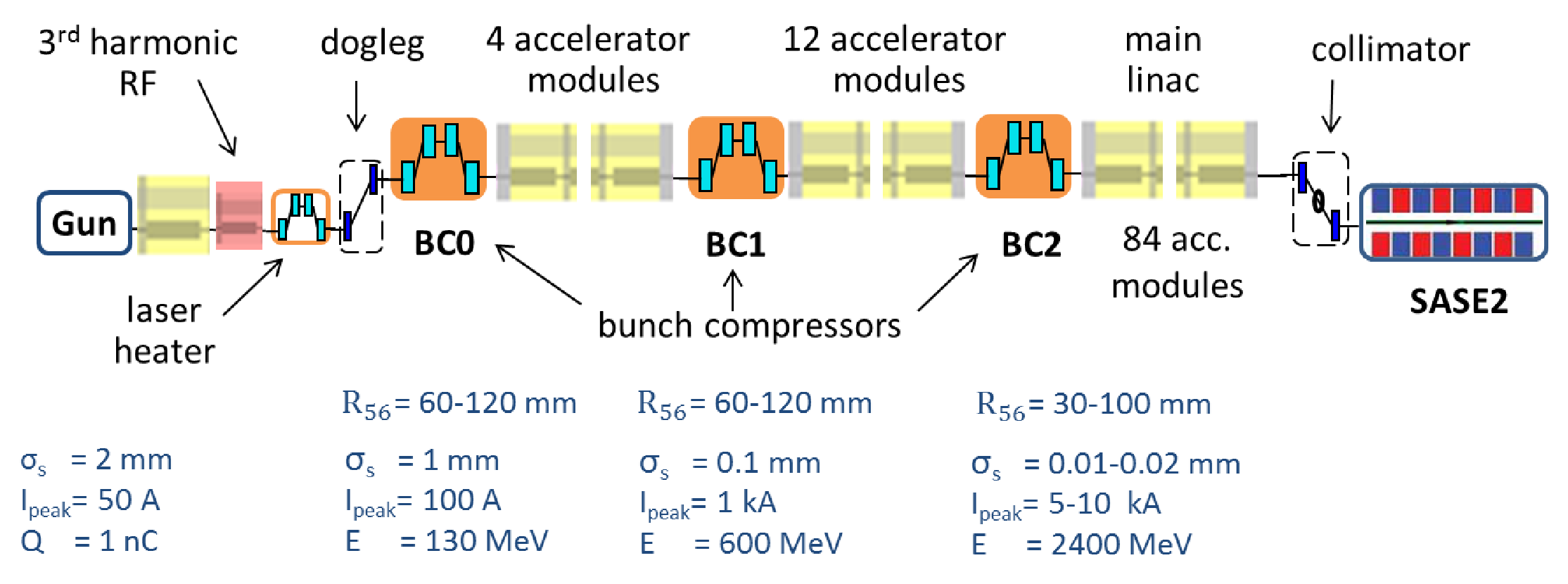}
\caption{Schematic layout of the European XFEL accelerator providing
  electrons up to 17.5 GeV electron energy in a macro pulse pattern
  with 27000 pulses/s. Details of the HXRSS SASE-2 undulator are shown
  in Fig.~\ref{layout2}.} \label{layout1}
\end{figure}

\begin{figure}
\includegraphics[width=0.5\textwidth]{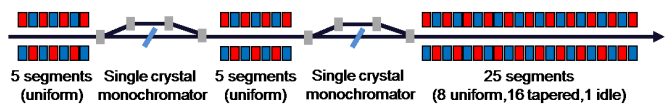}
\caption{Layout of SASE-2 undulator (35 segments) in the
  double-cascade self-seeding scheme for HXRSS. The monochromators are
  placed in the photon beam in between undulator segments where a
  magnetic chicane deviates the electrons.} \label{layout2}
\end{figure}

This section describes a configuration of the SASE-2 X-ray source at
the European XFEL, combining hard X-ray self-seeding (HXRSS) and
undulator tapering techniques in order to optimize the average output
spectral flux around $9$ keV, which is the optimum working point of
the UHRIX setup. In its simplest configuration, a HXRSS setup consists
of an input undulator and an output undulator separated by a chicane
with a single crystal monochromator \cite{GKS11}. Like this, it has
been implemented both at LCLS \cite{HXRSS12} and at SACLA
\cite{ITA14}. The time structure of the European XFEL is characterized
by $10$ macropulses per second, each macropulse consisting of $2700$
pulses, with $4.5$ MHz repetition rate inside the macropulse. The
energy carried by each pulse and the performance of the crystal
cooling system, removing deposited heat between macropulses, should
conservatively satisfy the condition that during a macropulse, the
drift in the central frequency of the crystal transmission function
cannot exceed the Darwin width. Then, due to the high repetition rate
of the European XFEL, the simplest two-undulator configuration for
HXRSS is not optimal and a setup with three undulators separated by
two chicanes with monochromators is proposed. This
amplification-monochromatization double cascade scheme is
characterized by a small heat load on the crystals and a high spectral
purity of the output radiation \cite{TWOC} \footnote{After successful
  demonstration of the self-seeding setup with a single crystal
  monochromator at the LCLS, it was decided that a double-cascade
  self-seeding scheme should be enabled at the SASE-2 beamline of the
  European XFEL from an early stage of operation \cite{SXSEED}.}.

The figure of merit to optimize for IXS experiments is the average
spectral photon flux. Here, the high-repetition rate of the European
XFEL yields a clear advantage compared with other XFELs. However, even
relying on its high repetition rate, the maximum output of the
European XFEL is $10^{12}~\mathrm{{ph}}/\mathrm{s}/\mathrm{meV}$ in
SASE mode at saturation, which is too low to satisfy the flux
requirements discussed in the previous section. Therefore self-seeding
and undulator tapering are needed.

The techniques proposed in this article exploit another unique feature
of the European XFEL, namely its very long undulators. The SASE-2 line
will feature $35$ segments, each consisting of a $5$ m long undulator
with $40$ mm period. The 175~m SASE-2, undulator is much longer than
required to reach saturation at $9$ keV (at $17.5$ GeV electron energy
and $250$ pC pulse charge the saturation length amounts to about $60$
m). We exploit this additional length to operate the SASE-2 baseline
in HXRSS mode followed by post-saturation tapering according to the
scheme in Fig. \ref{layout2}, which has been optimized for our
purposes.

\begin{table}[b!]
\caption{Operation parameters of the European XFEL used in this paper.}
\centering
\begin{tabular}{ l c c}
\hline & ~  &  ~Units \\ \hline
Undulator period      & 40                  & mm     \\
Periods per segment      & 125                   & -   \\
Total number of segments & 35                   & -    \\
K parameter (rms)     & 2.658                   & -  \\
Intersection length   & 1.1                   & m   \\
Wavelength            & 0.1358                  & nm \\
Energy                & 17.5                 & GeV    \\
Charge                & 250                  & pC\\
\hline
\end{tabular}
\label{tt1}
\end{table}

\begin{figure*}
\includegraphics[width=0.90\textwidth]{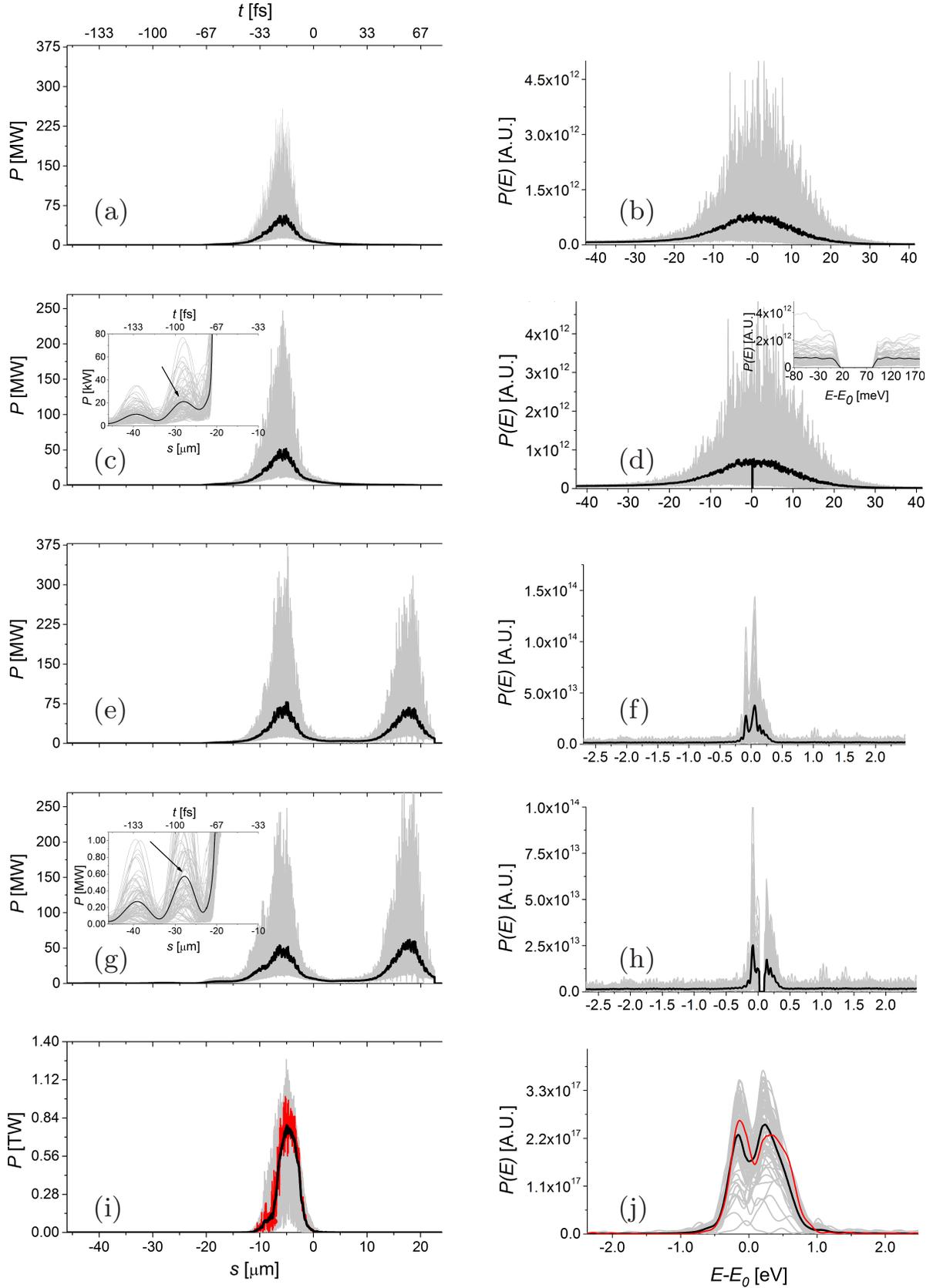}
\caption{Power distribution and spectrum of the X-ray pulse along the
  undulator: (a) and (b) calculated at the exit of the first undulator
  (5 segments); (c) and (d) after the first HXRSS monochromator; (e)
  and (f) at the exit of the second undulator (5 segments); (g) and
  (h) after the second HXRSS monochromator; (i) and (j) at the exit of
  the setup. Grey lines refer to single shot realizations, the black
  line refers to the average over one hundred simulations. The insets
  in (c) and (g) show an enlarged portion of the main plot,
  illustrating the seed appearing after the filtering process. The
  black arrows indicate the position of the seed relative to the
  electron slice with maximum current. The red lines in graphs (i) and
  (j) refer to the particular XFEL pulse that is used for wavefront
  propagation simulations (see Section~\ref{xrayoptics}). }
\label{pande}
\end{figure*}

As discussed above, since we seek to combine the high repetition rate
of the European XFEL with the HXRSS mode of operation, special care
must be taken to ensure that the heat load on the crystal does not
result in a drift in the central frequency of the transmission
function of more than a Darwin width. A preliminary estimate
\cite{HSINN} showed that in the case of radiation pulses with an
energy of a few $\mu$J, the heat deposited could be removed by the
monochromator cooling system without any problems \footnote{More
  precisely, that study considered X-ray pulses of $3~\mu$J, with a
  transverse size of $35~\mu$m FWHM, an energy of $8.2$ keV at a
  repetition rate of $4.5$ MHz. In that case, the drift of the central
  frequency for $1000$ pulses is within the Darwin width of
  reflection.}. In order to keep the pulse energy impinging on the
crystal within the few-$\mu$J range, one can exploit the
double-cascade self-seeding setup in Fig. \ref{layout2}. The setup
increases the signal-to-noise ratio, the signal being the seed pulse,
competing with the electron beam shot noise. At the position of the
second crystal, the seed signal is characterized by a much narrower
bandwidth than the competing SASE signal leading to a much higher
spectral density. In other words, in the frequency domain, the seed
signal level is amplified with respect to the SASE signal by a factor
roughly equal to the ratio between the SASE bandwidth and the seed
bandwidth. One can take advantage of the increased signal-to-noise
figure to reduce the number of segments in the first and second part
of the undulator down to five, thus reducing heat load on the crystals
due to impinging X-ray pulses. In the simulations we assume that the
diamond crystal parameters and the (004) Bragg reflection are similar
to those used for self-seeding at LCLS \cite{HXRSS12}. Optimization of
crystal thickness and the choice of reflections may yield an increase
in the final throughput \cite{YS13}. However, here we will not be
concerned with the optimization of the HXRSS setup in this respect.

\subsection{\label{sec:radi} Radiation from the SASE-2 undulator}

We performed numerical simulations of the high average-flux source in
Fig. \ref{layout2} using the GENESIS code \cite{GENESIS}. Simulations
are based on a statistical analysis consisting of $100$
runs. Start-to-end simulations \cite{S2ER} yielded information about
the electron beam (not shown) that is used as input for GENESIS. The
parameters pertaining to the double-cascade self-seeded operation mode
studied in this paper are shown in Table \ref{tt1}. The first five
undulator segments serve as a SASE radiator yielding the output power
and spectrum shown in Fig.~\ref{pande}(a) and (b), respectively.  As
explained in the previous Section, when working at high repetition
rates it is critical to minimize the energy per pulse impinging on the
diamond crystals. The energy per pulse can easily be evaluated
integrating the power distribution in Fig.~\ref{pande}(a) yielding an
average of about $1.2~\mu$J per pulse. As discussed in the previous
Section, this level of energy per pulse is fully consistent with the
proposed setup. The filtering process performed by the first crystal
is illustrated in Fig.~\ref{pande}(c) and (d).  The X-ray pulse then
proceeds through the second undulator as shown in Fig. \ref{layout2},
where it seeds the electron beam.

Power and spectrum at the exit of the second undulator are shown in
Fig.~\ref{pande}(e) and (f), respectively.  This figure illustrates
the competition between seed amplification and the SASE process, given
the relatively low seeded pulse power from the first part of the
setup. This is particularly evident in the time domain, where the
seeded pulse follows about $20~\mu$m after the SASE pulse with almost
similar power levels. Moreover, each of the pulses (seeded and SASE)
carries about the same energy as the initial SASE pulse incident on
the first crystal with a total incident average energy per pulse of
about $2.7~\mu$J, {\em i.e.} still within the heat-load limits
discussed in the previous Section. In the frequency domain a greatly
increased peak power spectral density is observed for the seeded
signal (compare Fig.~\ref{pande}(d) and (f)) while the SASE pulse
contributes a wide-bandwidth, noisy background. The fact that the
power spectral density for the seed signal is larger than for SASE by
about an order of magnitude (roughly corresponding to the ratio of the
SASE bandwidth to the seeded bandwidth) is what actually allows the
X-ray beam to impinge on the second HXRSS crystal at low power, but
with a large signal-to-noise (seeded-to-SASE) ratio, thus reducing
heat loading effects by about one order of magnitude compared to a
single-chicane scheme.

The filtering process performed by the second crystal is illustrated
in Fig.~\ref{pande}(g) and (h), respectively.  After this, the seed
signal is amplified to saturation and beyond, exploiting a combination
of HXRSS with post-saturation tapering.

\begin{figure}
\includegraphics[width=0.40\textwidth]{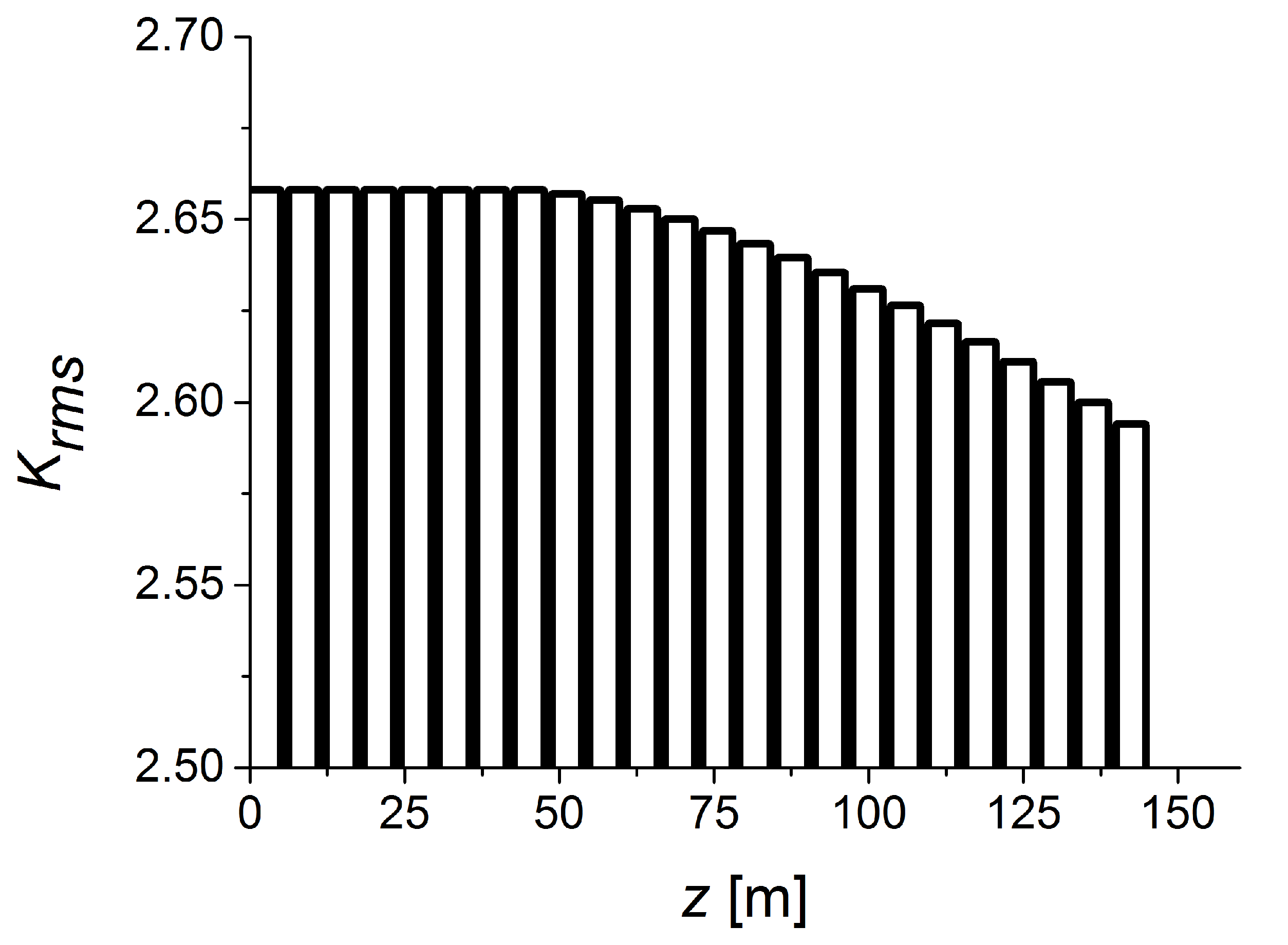}
\caption{Taper configuration for the output undulator (25 segments: 8
  uniform, 16 tapered, 1 idle). } \label{Taplaw}
\end{figure}

\begin{figure*}
\includegraphics[width=0.90\textwidth]{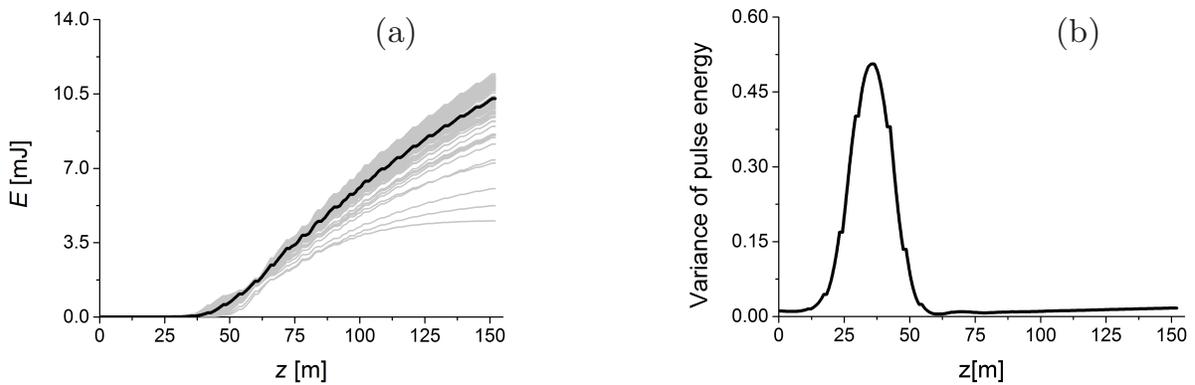}
\caption{Energy (a) and variance (b) of energy fluctuations of the
  seeded FEL pulse as a function of the distance inside the output
  undulator. Grey lines refer to single shot realizations, the black
  line refers to the average over a hundred
  realizations.} \label{enevar}
\end{figure*}

\begin{figure*}
\includegraphics[width=0.90\textwidth]{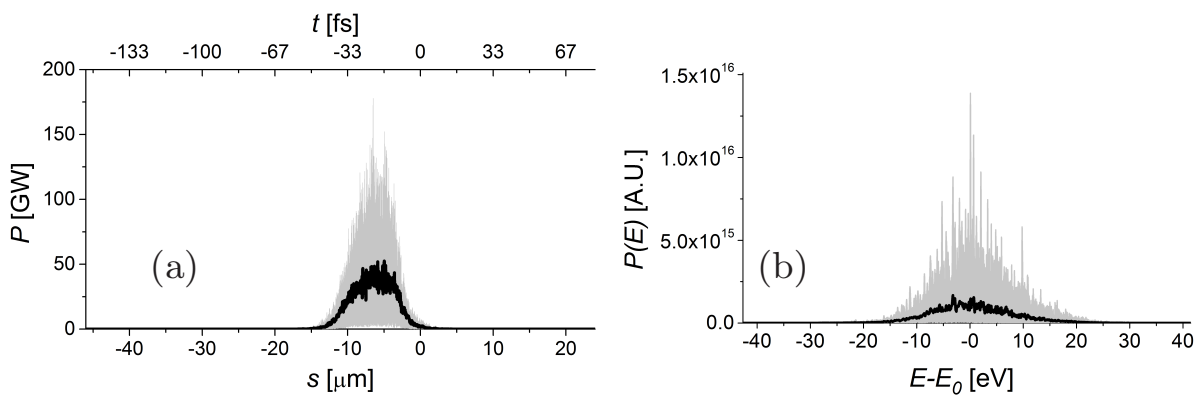}
\caption{Power (a) and spectrum (b) in the conventional SASE mode of
  operation at saturation, to be compared with power and spectrum in
  the HXRSS mode in Fig.~\ref{pande}(i) and (j), respectively. Grey
  lines refer to single shot realizations, the black line refers to
  the average over a hundred realizations.} \label{sasesat}
\end{figure*}

Tapering is implemented by changing the $K$ parameter of the
undulator, segment by segment according to Fig. \ref{Taplaw}. The
tapering law used in this work has been implemented on an empirical
basis, in order to optimize the spectral density of the output
signal. The use of tapering together with monochromatic radiation is
particularly effective, since the electron beam does not experience
brisk changes of the ponderomotive potential during the slippage
process.

\begin{figure*}[t!]
\includegraphics[width=0.8\textwidth]{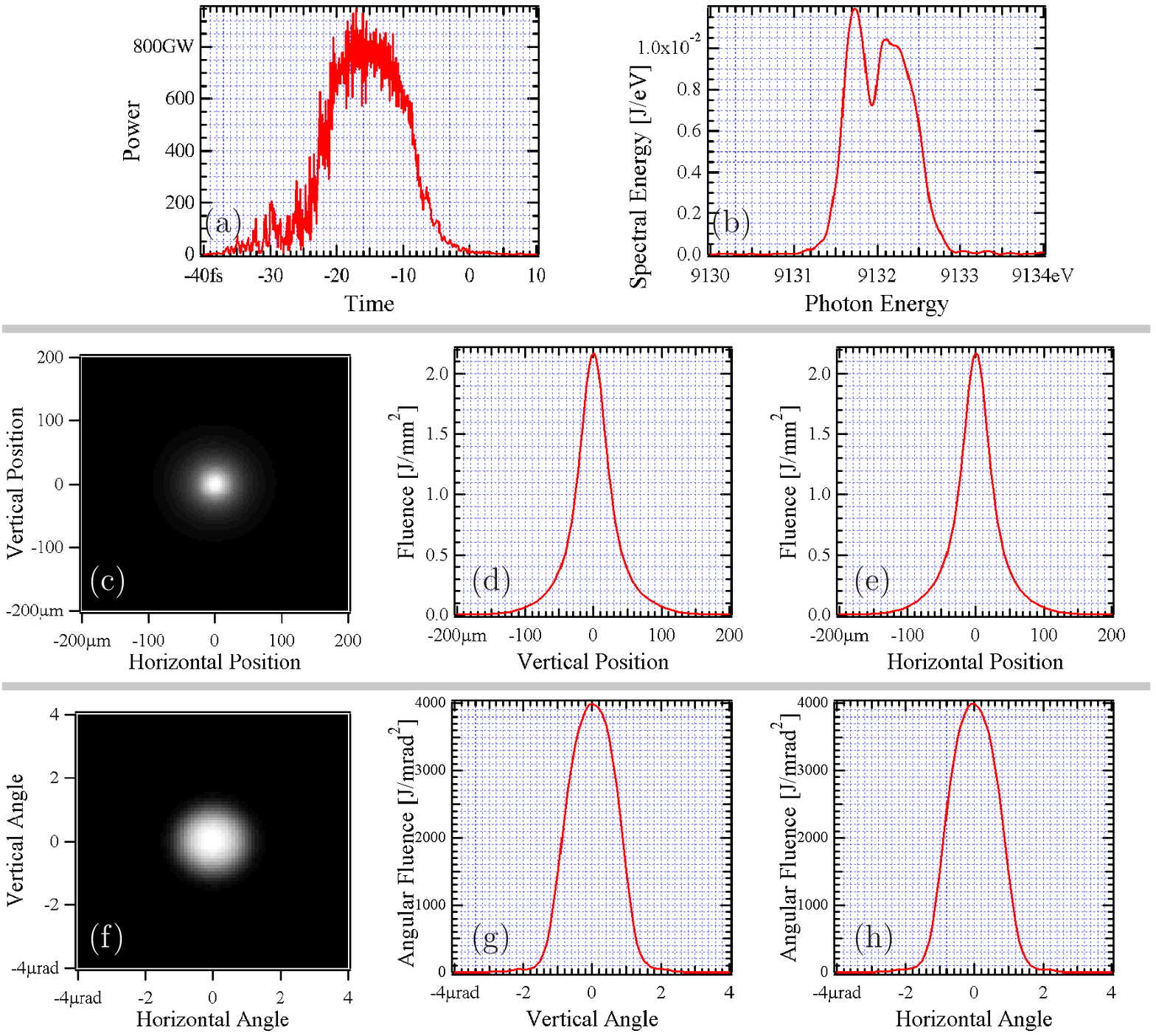}
\caption{Temporal, spectral, spatial, and angular distributions of the
  radiation pulse at the FEL undulator exit ($z=74$~m in
  Fig.~\ref{fig001}).  (a) Pulse power, pulse duration is $\simeq
  14$~fs (FWHM) (b) Spectrum, spectral bandwidth is $\simeq 0.95$~eV
  (FWHM). (c) Spatial distribution, 2D plot; (d) vertical cut through
  the center of the fluence distribution; and (e) horizontal cut. The
  beam size is about $50~\mu$m~(V)$\times 50~\mu$m~(H) (FWHM).  (f)
  Angular distribution, 2D plot; (g) vertical cut through the center
  of the fluence distribution; and (h) horizontal cut. The beam
  divergence amounts to $1.8~\mu$rad~(V)$\times 1.8~\mu$rad~(H)
  (FWHM). }
\label{fig00999}
\end{figure*}

The energy and variance of energy fluctuations of the seeded FEL pulse
as a function of the distance inside the output undulator are
illustrated in Fig. \ref{enevar}. On the average, pulses of about
11~mJ energy can be produced with this scheme. The final output of our
setup is presented in Fig.~\ref{pande}(i) and (j), respectively, in
terms of power and spectrum. This result should be compared with the
output power and spectrum for SASE at saturation in Fig. \ref{sasesat}
corresponding to the {\em conventional} operation mode foreseen at the
European XFEL. Considering an average over $100$ shots, the peak power
for the SASE saturation case in Fig. \ref{sasesat} is about $4 \times
10^{10}$ W, while for the seeded case in Fig.~\ref{pande}(i), it has
grown to $7.5 \times 10^{11}$ W. This corresponds to an increase in
flux from about $7 \times 10^{11}$ photons per pulse to about $7
\times 10^{12}$ photons per pulse. This amplification of about one
order of magnitude is due to tapering. In addition, the final SASE
spectrum has a FWHM of about $11.6$ eV, corresponding to a relative
bandwidth of $1.2 \times 10^{-3}$ while, due to the enhancement of
longitudinal coherence, the seeded spectrum has a FWHM of about
0.94~eV, corresponding to a relative bandwidth of $1 \times 10^{-4}$.

In conclusion, the proposed double-cascade self-seeding tapered scheme yields one order of magnitude increase in peak power due to undulator tapering, and a bit less than an order of magnitude decrease in spectral width due to seeding. Combining the two effects, we obtain an increase in spectral flux density of more that two orders of magnitude compared to saturated SASE ($2.1 \times 10^{14}~\mathrm{{ph}}/\mathrm{s}/\mathrm{meV}$ compared to $1.5 \times 10^{12}~\mathrm{{ph}}/\mathrm{s}/\mathrm{meV}$). The transverse beam size and divergence at the exit of the undulator are shown in Figs.~\ref{fig00999}(c)-(e) and \ref{fig00999}(f)-(h), respectively. The beam profile is nearly circular with a size of about $50~\mu$m (FWHM) and a divergence of about $1.8~\mu$rad (FWHM). In the next section we will complement this information with detailed wavefront propagation simulations through the optical transport line up to the UHRIX setup.

\section{Optics for Ultra-High-Resolution IXS}
\label{xrayoptics}

The desired ultra-high-resolution IXS studies with 0.1~meV spectral and 0.02~nm$^{-1}$ momentum transfer resolution require a significant amount of X-ray photons with energy $E_{\ind{0}}=9.13185$~keV and momentum $K = E_{\ind{0}}/\hbar c = 46.27598$~nm$^{-1}$ to be delivered to the sample within $\Delta E \lesssim 0.1$~meV spectral bandwidth and a transverse momentum spread $\Delta K \lesssim 0.02$~nm$^{-1}$, all concentrated on the sample in a spot of $\Delta s \lesssim 5~\mu$m (FWHM) diameter. The aforementioned photon energy $E_{\ind{0}}$ is fixed by the (008) Bragg reflection from Si single crystals, one of the central components of the ultra-high-resolution optics presented in detail below.

\begin{figure*}[t!]
\includegraphics[width=1.0\textwidth]{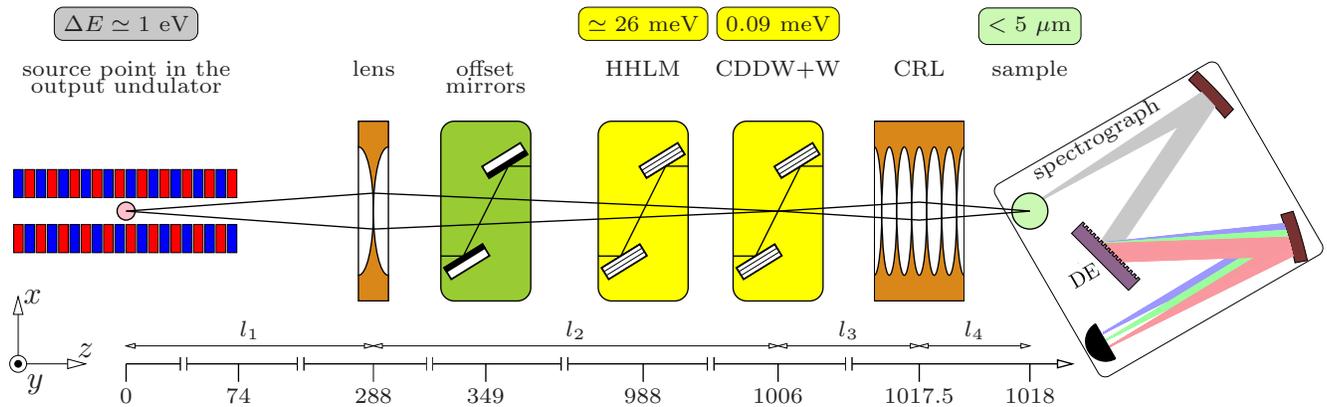}
\caption{Main optical components of the proposed UHRIX instrument at
  the SASE-2-undulator beamline of the European XFEL shown
  schematically together with the output undulator. Optical components
  are presented as pictographs positioned at various distances from
  the effective source position in the SASE-2 undulator, 74~m upstream
  of the undulator exit. See text for descriptions.}
\label{fig001}
\end{figure*}

We consider a scenario in which the UHRIX instrument is installed at
the SASE-2-undulator beamline of the European XFEL. In particular, we
consider an option of integrating UHRIX into the Materials Imaging and
Dynamics (MID) station \cite{MID13}, an instrument presently under
construction at the European XFEL. A schematic view of the optical
components essential for delivering photons with the required
properties to the sample is shown in Fig.~\ref{fig001}. Optics are
shown as pictographs in certain distances from the source. The
effective source position is located around 74 m inside the undulator
measured from the exit. This number was determined by back-propagation
in free space of the XFEL radiation from the undulator end.

The main optical components are as follows: A biconcave parabolic
refractive lens \cite{LST99}, creates a secondary source on the
6-bounce angular dispersive ultra-high-resolution CDDW+W
monochromator. This is essential in order to achieve a tight focal
spot on the sample because it eliminates the blurring that the strong
angular dispersion of the CDDW+W monochromator would cause otherwise
\cite{Shvydko15}. The CDDW+W monochromator then selects a 0.1~meV
spectral bandwidth from the incident X-ray beam. The CDDW+W is a
modification of a CDW-type angular dispersive monochromator~
\cite{SLK06,ShSS11,SSS13}, which uses a three-step process of
collimation (C), angular dispersion (D), and wavelength selection
(W)~\cite{Shvydko-SB}.  Finally, a parabolic compound refractive lens
\crls \cite{SKSL,LST99} focuses the monochromatic X-rays on the
sample.

The X-ray spectrograph captures photons scattered from the source in a
sufficiently large solid angle and images them in a few-meV wide
spectral window with 0.1-meV spectral resolution in the dispersion
plane. The dispersing element (DE), a hard X-ray analog of an optical
diffraction gratings, is a key component of the spectrograph. The
spectrograph is also capable of simultaneously imaging scattered
intensity perpendicular to the dispersion plane in a range of
0.2-nm$^{-1}$ with 0.01-nm$^{-1}$ resolution. Supplementary optical
components include a pair of offset mirrors ($z$=349~m) which separate
the beam from unwanted high-energy bremsstrahlung, and the two-bounce,
two-crystal non-dispersive high-heat-load monochromator (HHLM at
$z$=988~m). The HHLM narrows the 1~eV bandwidth of the incident X-rays
to about 26~meV and thus reduces the heat-load onto the CDDW+W
monochromator by a factor of 36.

In the remaining parts of this section, the choice of optical elements
is justified and their design parameters are determined, first by
using dynamical theory calculations for monochromatization with the
X-ray crystal optics components in Section~\ref{monos} and then by
applying ray transfer matrix formalism for ray tracing in
Section~\ref{ray}. The optical design is verified by wavefront
propagation simulations using a combined application of GENESIS
\cite{GENESIS} and SRW \cite{SRW} codes with results presented in
Section~\ref{wave}.

\subsection{Monochromatization of X-rays}
\label{monos}

The radiation from the undulator discussed previously has about
950~meV bandwidth. It must be reduced to 0.1~meV and delivered to the
sample with the smallest possible losses. To this end the previously
discussed HHLM and CDDW+W are used in a two-tiered monochromatization
scheme. In the following subsections we discuss their operating
principles and design parameters in detail.

\begin{figure}[t!]
\setlength{\unitlength}{\textwidth}
\includegraphics[width=0.4\textwidth]{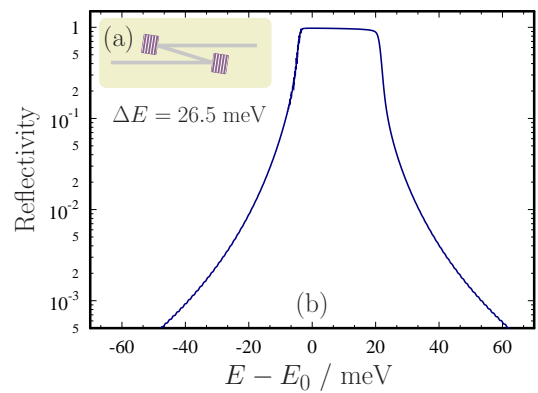}
\caption{(a) Schematic view of the high-heat-load monochromator
  (HHLM). (b) Dynamical theory calculations of the spectral
  distribution of x~rays around the nominal photon energy $E_{\ind{o}}
  = 9.13185$~keV after two successive (115) Bragg reflections from
  diamond. The spectral bandwidth of the transmitted X-rays is
  26.5~meV with a peak reflectivity of 97.7\%. The angular spread of
  the incident X-rays is $\Delta\theta_{\indrm{inc}}=1~\mu$rad.}
\label{fig002}
\end{figure}

\begin{table}[t!]
\centering
\begin{tabular}{|l|lllllll|}
  \hline
crystal/ &$\vc{H}$  &$\eta_{\ind{H}} $  & $\theta_{\ind{H}} $  &  $d$ &  $b_{\ind{H}}$ & $\dei{H} $ &  $\dai{H}$  \\[-5pt]
function &          &                                  &                 &       &              &            &        \\
         & $(hkl)$  & [deg]                    &  [deg]         & [mm]   &              & [meV]      &  [$\mu$rad] \\[-5pt]
         &          &                                  &                &        &              &            &        \\[0pt]
  \hline
C* / 1st & (1 1 5) & 0 &  81.45 & 0.1 & -1 & 33  & 24   \\[0pt]
  \hline
C* / 2nd & (1 1 5)  & 0 &  81.45 & 0.3 & -1 & 33  & 24  \\[0pt]
  \hline
\end{tabular}
\caption{Crystal and Bragg reflection parameters of the crystal elements of the HHL monochromator: $(hkl)$ - Miller indices of the Bragg diffraction vector $\vc{H}$; $\eta_{\ind{H}}$ - asymmetry angle; $\theta_{\ind{H}}$ - glancing angle of incidence; $d$ - crystal thickness; $b_{\ind{H}}=-\sin(\theta_{\ind{H}}+\eta_{\ind{H}})/\sin(\theta_{\ind{H}}-\eta_{\ind{H}})$ - asymmetry parameter;  $\dei{H}$ and $\dai{H}$ are the Bragg reflection's intrinsic spectral width, and angular acceptance, respectively.}
\label{tab2}
\end{table}

\subsubsection{High-heat-load monochromator }

A schematic of the high-heat-load monochromator (HHLM) is shown in Fig.~\ref{fig002}(a).  In the present design two diamond (C*) crystal
plates are used as Bragg reflectors, with the (115) planes parallel to the crystal surface (symmetric Bragg). The (115) reflection is chosen for the Bragg angle to be as close as possible to $90^{\circ}$ (backscattering) for 9.13185~keV X-rays. This is dictated by stability requirements under high heat load, as the spectral variation of the reflected X-rays with incidence angle is minimized in back-scattering geometry. The Bragg reflection and crystal parameters used in the HHLM are provided in Table~\ref{tab2}. Dynamical theory calculations of the spectral
distribution of X-rays around the nominal photon energy $E_{\ind{o}} = 9.13185$~keV after two successive (115) Bragg reflections from diamond
are shown in Fig.~\ref{fig002}(b).

\subsubsection{High-resolution monochromator CDDW+W}

\begin{figure}[t!]
\setlength{\unitlength}{\textwidth}
\includegraphics[width=0.5\textwidth]{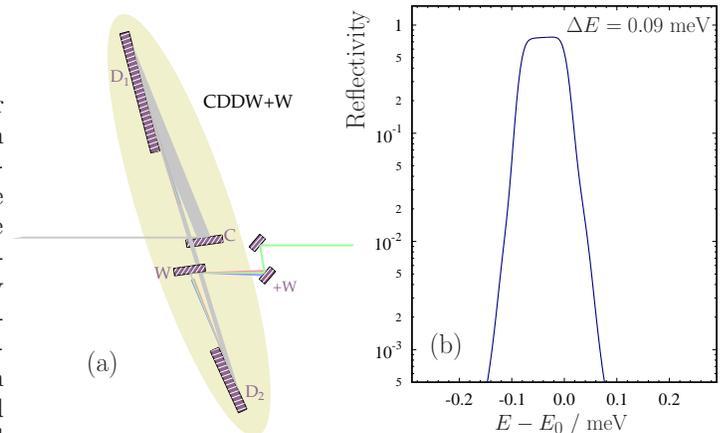}
\caption{(a) Schematic view of the CDDW+W monochromator. (b) Dynamical
  theory calculations of the spectral distribution of x~rays after six
  successive reflection from the crystals of the CDDW+W optic.
  Calculations were performed for incident X-rays around the nominal
  photon energy $E_{\ind{0}} = 9.13185$~keV, with an angular spread of
  $1~\mu$rad and crystal parameters as in Table~\ref{tab1}. The peak
  reflectivity of the optic is 71\% with a spectral bandwidth of
  0.09~meV.}
\label{fig003}
\end{figure}

The CDDW+W monochromator is a modification of the CDDW monochromator
\cite{ShSS11,SSS13,SSS14} complemented by two additional
wavelength-selector crystals +W, ensuring a substantially reduced
bandwidth and sharp Gaussian tails in the resolution function
\cite{Shv11,Shvydko12,SSM13}. Figure~\ref{fig003}(a) shows a schematic
view of the CDDW+W monochromator, while Fig.~\ref{fig003}(b) presents
the results of dynamical theory calculations of the spectral
distribution of X-rays after the CDDW+W. The crystal parameters used
in the calculations are given in Table~\ref{tab1}. The nominal photon
energy $E_{\ind{0}}$=9.13185~keV of the UHRIX instrument is determined
by the (008) Bragg reflection from the Si dispersion crystals D$_1$
and D$_2$ with a Bragg angle of $\theta=89.5^{\circ}$.

\begin{table}[b!]
\centering
\begin{tabular}{|l|llllllll|}
  \hline
crystal/ &$\vc{H}$ & $\eta_{\ind{H}} $ &  $\theta_{\ind{H}} $  &  $d$ &  $b_{\ind{H}}$ & $\dei{H} $ &  $\dai{H}$   & $\dirate_{\ind{H}}$ \\[-5pt]
function &                        &                  &                 &       &              &            &              &      \\
         & $(hkl)$       & [deg]            &  [deg]         & [mm]   &              & [meV]      &  [$\mu$rad]  & [$\mu$rad \\[-5pt]
         &                        &                  &                &        &              &            &              &  meV]   \\[0pt]
  \hline
C* / C & (3 3 1 )& -48  & 56.06 & 0.5 & -0.14 & 124 & 20   & -0.1 \\[0pt]
  \hline
Si /D$_{\ind{1}}$ & (8 0 0 ) & 87.5  &  89.5 & 10  & -1.5 &   22 & 280 & 6.2 \\[0pt]
  \hline
Si /D$_{\ind{2}}$ & (8 0 0 ) & 87.5  &  89.5 & 10 & -1.5 &  22 & 280  &-6.2  \\[0pt]
  \hline
C* / W  & (3 3 1 ) & 48 &    56.05 & 0.5 & -6.9 &  18 & 2.9 & 0.9  \\[0pt]
  \hline
C* / +W  & (4 0 0 ) & 0 &  49.57 & 0.5 & -1.0 &  75  & 10 & 0  \\[0pt]
  \hline
C* / +W   & (4 0 0 ) &  0 &  49.57 & 0.5 & -1.0 &  75  & 10  & 0  \\[0pt]
  \hline
\end{tabular}
\caption{Elements of the CDDW+W optics with their crystal and Bragg reflection parameters. Similar definitions are used as in Table~\ref{tab2}. In addition, $\dirate_{\ind{H}}$ is the Bragg reflection's dispersion rate. The cumulative asymmetry parameter and dispersion rate of the  monochromator are $\dcomm{6}=2.25$, and $\fcomm{6}=112~\mu$rad/meV, see definition in Table~\ref{tab3}.
  The X-ray photon energy is $E_{\ind{0}}=9.13185$~keV.}
\label{tab1}
\end{table}

\subsection{Focusing optics}
\label{focusingoptics}

Because of the very large distances $l_{\ind{1}}$ and $l_{\ind{2}}$ a single 2D parabolic Be lens \cite{LST99}, denoted in Fig.~\ref{fig001}
as ``\crl'', is sufficient to focus X-rays onto the CDDW+W monochromator. A lens with 1.68 mm radius ($R$) at the parabola apex, a focal distance $f_{\mathrm {\crl}}=R/2\delta=205.5$~m, and with 1.5-mm geometrical aperture is considered in the following. The corrections $\delta = 4.08684 \times 10^{-6}$ and $\beta=1.4201 \times 10^{-9}$ to the refractive index $n=1-\delta-{\mathrm i}\beta$ \cite{HGD93} are used in the wavefront propagation calculations.

The \crls\ at $z$=1017.5~m, see Fig.~\ref{fig001}, focuses X-rays from the secondary source at the CDDW+W monochromator onto the sample. In preliminary wavefront propagation simulations an idealized system will be considered consisting of $N=39$ lenses each of 152.75~$\mu$m radius $R$ and all placed at the same position. The total focal length of the lens assembly is $f_{\mathrm{\crl}}=R/2N\delta=0.479$~m. In the final calculations a more realistic extended \crls\ will be used containing 41 individual lenses separated by a 3~mm distance, with the first 39 having a 150~$\mu$m radius, and the last two a 400~$\mu$m radius at the parabola apex. The geometrical aperture of the \crls\ is 1~mm, which does not truncate the incident wavefront. All lenses are assumed to be perfect.

\subsection{Focal spot size and momentum spread on the sample - analytical ray tracing}
\label{ray}

We use the ray-transfer matrix technique \cite{KL66,MK80-1,Siegman} to
propagate paraxial X-rays through the optical system of the UHRIX
instrument and to determine linear and angular sizes of the X-ray beam
along the optical system. In a standard treatment, a paraxial ray in
any reference plane (a plane perpendicular to the optical axis $z$) is
characterized by its distance $x$ from the optical axis, by its angle
$\xi$ with respect to that axis, and the deviation $\delta E$ of the
photon energy from a nominal value $E$.  The ray vector
$\vc{r}_{\ind{1}}=(x,\xi,\delta E)$ at an input reference plane
(source plane) is transformed to
$\vc{r}_{\ind{2}}=\hat{O}\vc{r}_{\ind{1}}$ at the output reference
plane (image plane), where $\hat{O}=\{ABG,CDF,001\}$ is a ray-transfer
matrix of an optical element (elements) placed between the planes. The
upper rows of Table~\ref{tab3} present the ray transfer matrices of
the major components of the UHRIX optical system. The ray transfer
matrix $\hat{\uhrix}$ of the UHRIX instrument, which describes
propagation from the source to the sample, is presented in the last
row of Table~\ref{tab3}. We refer to Ref.~\cite{Shvydko15} for details
about the derivation of these matrices and provide here only essential
notation and definitions.

In the focusing system, see the matrix
$\hat{\focus}(l_{\ind{2}},f,l_{\ind{1}})$ in Table~\ref{tab3}, a
source in a reference plane at a distance $l_{\ind{1}}$ upstream of a
lens with focal length $f_{\ind{12}}$ is imaged onto the reference
image plane located at a distance $l_{\ind{2}}$ downstream from the
lens. If the parameter $\Delta_{\ind{12}}$ defined in Table~\ref{tab3}
equals zero, the classical lens equation
${l_{\ind{1}}}^{-1}+{l_{\ind{2}}}^{-1}={f_{\ind{12}}}^{-1}$ holds. In
this case, the system images the source with inversion and a
magnification factor
$\mu_{\ind{2}}=1/\mu_{\ind{1}}=-l_{\ind{2}}/l_{\ind{1}}$ independent
of the angular spread of rays in the source plane.

In the ray transfer matrix $\hat{\crystal}(b,\sgn \dirate)$ describing
Bragg reflection from a crystal at angle $\theta$, the asymmetry
factor $b$ determines how the beam size and divergence change upon
Bragg reflection. The angular dispersion rate $\dirate$ describes how
the photon energy variation $\delta E$ from a nominal value $E$
changes the reflection angle with a fixed incident angle. The Bragg
reflecting atomic planes are assumed to be at an asymmetry angle
$\eta$ with respect to the crystal surface.

The ray transfer matrix $\hat{\crystal}_{n}(\dcomm{n},\fcomm{n})$
describing successive Bragg reflections from a system of $n$ crystals,
has the same structure as that of a single Bragg reflection. The only
difference is that the asymmetry parameter $b$ and the angular
dispersion rate $\dirate$ are substituted by the appropriate
cumulative values $\dcomm{n}$ and $\fcomm{n}$, respectively. The ray
transfer matrices of the offset mirrors and of the HHLM consisting of
two symmetric Bragg reflections ($\eta=0$, $b=-1$, $\dirate=0$) (see
Table~\ref{tab2}) are unit matrices, leading to no change in the beam
parameters.

\begin{table*}[t!]
\centering
\begin{tabular}{|l|l|l|}
  \hline
& &  \\
  Optical system & Ray-transfer matrix $\{ABG,CDF,001\}$ & Definitions and Remarks\\[-5pt]
& &  \\
  \hline  \hline
& &  \\[-3.2mm]
\parbox[c]{0.25\textwidth}{Focusing system \\ \includegraphics[width=0.25\textwidth]{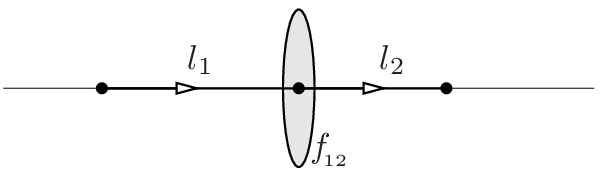}}  &
\parbox[c]{0.45\textwidth}{$\hat{\focus}(l_2,f_{\ind{12}},l_1) = $\\
$\left( \begin{array}{ccc} \mu_{\ind{2}} & \Delta_{\ind{12}} & 0  \\ -\frac{1}{f_{\ind{12}}}  & \mu_{\ind{1}} & 0\\ 0 & 0 & 1  \end{array} \right)$}  &
\parbox[c]{0.25\textwidth}{$\Delta_{\ind{12}}=l_{\ind{1}}+l_{\ind{2}}-\frac{l_{\ind{1}}l_{\ind{2}}}{f_{\ind{12}}}$\\$\mu_{\ind{2}}=1-\frac{l_{\ind{2}}}{f_{\ind{12}}}=-\frac{l_{\ind{2}}-\Delta_{\ind{12}}}{l_{\ind{1}}}$\\$\mu_{\ind{1}}=1-\frac{l_{\ind{1}}}{f_{\ind{12}}}=-\frac{l_{\ind{1}}-\Delta_{\ind{12}}}{l_{\ind{2}}}$ }     \\[-0.5mm]
\hline
& &  \\[-3.2mm]
\parbox[c]{0.25\textwidth}{Bragg reflection from a crystal\\ \includegraphics[width=0.25\textwidth]{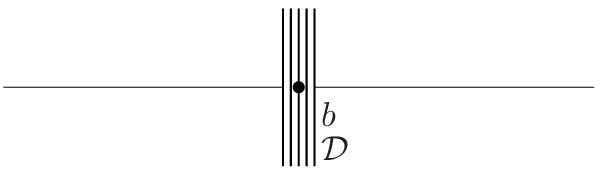}}   &
\parbox[c]{0.45\textwidth}{$\hat{\crystal}(b,\sgn \dirate) =$ \\ $\left( \begin{array}{ccc} {1}/{b} & 0 & 0  \\0  & b & \sgn\dirate \\ 0 & 0 & 1 \end{array} \right)$}  &
\parbox[c]{0.25\textwidth}{$b=-\frac{\sin(\theta+\eta)}{\sin(\theta-\eta)}$\\  $\dirate = -(1/E)(1+b)\tan\theta $\\ $\sgn = -1$ {\small for clockwise, and}  $\sgn = +1$ {\small for counterclockwise ray deflection}}     \\[-0.5mm]
\hline
& &  \\[-3.2mm]
\parbox[c]{0.25\textwidth}{Successive Bragg reflections from $n$  crystals  \\  \includegraphics[width=0.25\textwidth]{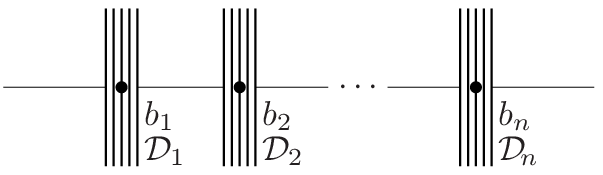}}  &
\parbox[c]{0.45\textwidth}{$\hat{\crystal}_{n}(\dcomm{n},\fcomm{n})=  \hat{\crystal}(b_n,\sgn_n\dirate_n)\dotsb   \hat{\crystal}(b_1,\sgn_1\dirate_1)$\\  $\left( \begin{array}{ccc} \acomm{n} & 0 & 0  \\ 0  & \dcomm{n}  & \fcomm{n} \\ 0 & 0 & 1 \end{array} \right)$}  &  \parbox[c]{0.25\textwidth}{$\dcomm{n}=b_1b_2b_3 \dotsc b_n$ \\ $\fcomm{n}=b_n\fcomm{n-1} + \sgn_n\dirate_n$\\ $\sgn_i=\pm 1$, $i=1,2,...,n$ }     \\[-0.5mm]
\hline
& &  \\[-3.2mm]
\parbox[c]{0.25\textwidth}{UHRIX \\ \includegraphics[width=0.25\textwidth]{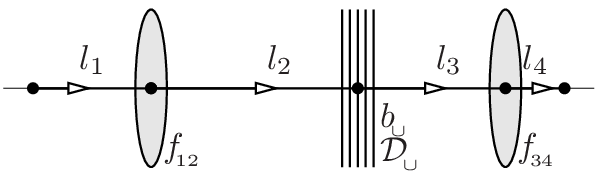}}   &
\parbox[c]{0.45\textwidth}{$ \hat{\uhrix} = \hat{\focus}(l_{\ind{4}},f_{\ind{34}},l_{\ind{3}}) \hat{\crystal}_{6}(\fcomm{6},\dcomm{6}) \hat{\focus}(l_{\ind{2}},f_{\ind{12}},l_{\ind{1}}) =$ \\
$ \left( \begin{array}{ccc} \frac{\mu_{\ind{2}}\mu_{\ind{4}}}{\dcomm{6}}-\frac{\dcomm{6} \Delta_{\ind{34}} }{f_{\ind{12}} }  & \frac{ \Delta_{\ind{12}}  \mu_{\ind{4}} }{\dcomm{6}} + \Delta_{\ind{34}}  \mu_{\ind{1}} \dcomm{6}   &  \Delta_{\ind{34}} \fcomm{6}  \\ -\frac{\mu_{\ind{2}}}{f_{\ind{34}} \dcomm{6}}-\frac{\mu_{\ind{3}} \dcomm{6}}{f_{\ind{12}}} & -\frac{ \Delta_{\ind{12}}}{\dcomm{6} f_{\ind{34}} } + \mu_{\ind{1}} \mu_{\ind{3}} \dcomm{6}   &  \mu_{\ind{3}} \fcomm{6}    \\ 0 & 0 & 1 \end{array}\!\!\!\!\! \right)
 $}  &  \parbox[c]{0.25\textwidth}{ }     \\
  \hline
\end{tabular}
\caption{Ray-transfer matrices for a focusing system, for Bragg reflection from crystals, and for the complete optical system of the UHRIX instrument from source to sample. }
\label{tab3}
\end{table*}

The total ray transfer matrix $\hat{\uhrix}$ of the UHRIX instrument
is a product of the ray transfer matrices of the \crl\ focusing system
$\hat{\focus}(l_{\ind{2}},f_{\ind{12}},l_{\ind{1}})$; the CDDW+W
six-crystal matrix $\hat{\crystal}_{6}(\dcomm{6},\fcomm{6})$, and of
the \crls\ focusing system
$\hat{\focus}(l_{\ind{4}},f_{\ind{34}},l_{\ind{3}})$. The asymmetry
parameters and the dispersion rate of the CDDW+W monochromator
crystals required for the CDDW+W matrix are provided in
Table~\ref{tab1}. $\hat{\uhrix}$ describes propagation of X-rays in
the vertical $(x,z)$ plane (see reference system in
Fig.~\ref{fig001}), in which the Bragg diffraction from the
monochromator crystals takes place. Propagation of X-rays in the
horizontal $(y,z)$ plane is not affected by Bragg diffraction from the
monochromator crystals. Here, the appropriate UHRIX ray-transfer
matrix is obtained from $\hat{\uhrix}$ with parameters $\dcomm{6}=1$
and $\fcomm{6}=0$.

To determine the actual focal size and angular spread on the sample we
use a linear source size (FWHM) $x_{\ind{0}} = y_{\ind{0}} =50~\mu$m,
and an angular source size $\xi_{\ind{0}}=1.8~\mu$rad, as derived from
the XFEL simulations in Section~\ref{sec:SASE-2}. The energy spread of
the X-rays is assumed $\delta E_{\ind{0}}=0.09$~meV. For the
cumulative asymmetry parameter and dispersion rate of the CDDW+W
monochromator we use $\dcomm{6}=2.25$, and $\fcomm{6}=112~\mu$rad/meV
as obtained from Table~\ref{tab1} and the distances between the
optical elements are $l_{\ind{1}}=288$~m, $l_{\ind{2}}=718$~m,
$l_{\ind{3}}=11.5$~m, and $l_{\ind{4}}=0.5$~m, see Fig.~\ref{fig001}.

\subsubsection{Focal spot size on the sample}
\label{focalspotsize}

The smallest focal spot size on the sample is achieved provided
$\Delta_{\ind{12}}=0$, that is, the \crl\ focuses X-rays on the CDDW+W
monochromator, and $\Delta_{\ind{34}}=0$, meaning that the \crls\
refocuses X-rays on the sample with the secondary source on the CDDW+W
monochromator. The focusing conditions require $f_{\ind{12}}=205.5$~m,
and $f_{\ind{34}}=0.479$~m for the focal distances for the \crl\ and
\crls , respectively, see also Sec.~\ref{focusingoptics}. In this
case, the elements $B$ and $G$ of the $\hat{\uhrix}$ matrix are zero
so the vertical and horizontal linear sizes of the source image on the
sample are determined only by the element $A$:
\begin{equation}
x_{\ind{4}}=   x_{\ind{0}} \mu_{\ind{2}}\mu_{\ind{4}}/\dcomm{6}, \hspace{0.5cm} y_{\ind{4}}=   y_{\ind{0}} \mu_{\ind{2}}\mu_{\ind{4}}.
\label{eq001}
\end{equation}
With $\mu_{\ind{2}}=-l_{\ind{2}}/l_{\ind{1}}=2.5$, and
$\mu_{\ind{4}}=-l_{\ind{4}}/l_{\ind{3}}=0.044$, we obtain for the
vertical spot size $x_{\ind{4}}=2.4~\mu$m, while for the horizontal
size $y_{\ind{4}}=5.4~\mu$m.  The vertical spot size $x_{\ind{4}}$ is
less than half the target specification (5~$\mu$m) required to achieve
0.1~meV spectral resolution of the spectrograph \cite{Shvydko15}, as
discussed below in Section~\ref{spectrograph}. If focusing onto the
CDDW+W is not perfect so that $\Delta_{\ind{12}}\not = 0$, this may
lead to an increase of the spot size by $\Delta
x_{\ind{4}}=\xi_{\ind{0}}{ \Delta_{\ind{12}}
  \mu_{\ind{4}}}/{\dcomm{n}} $ (resulting from element $B$ of the
UHRIX ray-transfer matrix).  However, this is not very critical, as
even with a mismatch of $\Delta_{\ind{12}} \simeq 10$~m, the spot size
increases only by an insignificant $\Delta x_{4}\simeq 0.4~\mu$m.

\begin{figure*}[t!]
\includegraphics[width=0.99\textwidth]{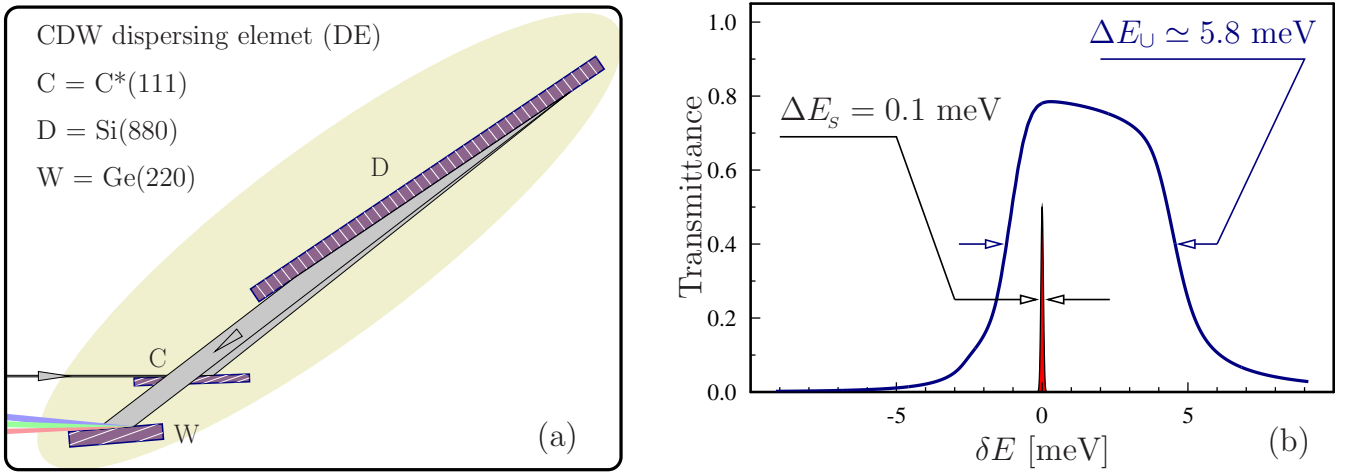}
\caption{(a) The CDW-type three-crystal dispersing element of the
  spectrograph. (b) The spectral transmission function of the
  spectrograph with the CDW dispersing element ensuring a 5.8-meV
  broad window of imaging. The sharp line presents an 0.1-meV design
  spectral resolution of the spectrograph.}
\label{fig008}
\end{figure*}

\subsubsection{Transverse momentum spread}

The transverse momentum spread in the diffraction plane (vertical) $\Delta K = K\xi_{\ind{4}}$ is defined by the angular spread
\begin{equation}
\xi_{\ind{4}}=\sqrt{ (C x_{\ind{0}})^2 + (D \xi_{\ind{0}})^2 + (F \delta E_{\ind{0}})^2}
\label{eq002}
\end{equation}
of X-rays incident on the sample\footnote{The beam sizes and the
  angular spread in Eqs.~\eqref{eq001}-\eqref{eq002} are obtained
  by propagation of second-order statistical moments, using transport
  matrices derived from the matrices presented in Table~\ref{tab3},
  and assuming zero cross-correlations (i.e. zero mixed second-order
  moments).}.
Here we assume a Gaussian distribution of the beam parameters. In the
vertical scattering plane the UHRIX ray-transfer matrix elements are
$C=2.56~\mu$rad/$\mu$m, $D=21$, and $F=-2.58~\mu$rad/$\mu$eV. With
$x_{\ind{0}}=50~\mu$m, $\xi_{\ind{0}}=1.8~\mu$rad, and $\delta
E_{\ind{0}}=90~\mu$eV we obtain $\xi_{\ind{4}}=265~\mu$rad, and
$\Delta K_{\ind{x}}=0.012$~nm$^{-1}$.

In the horizontal plane there is no angular dispersion. The cumulative
dispersion rate $\fcomm{6}=0$, and the asymmetry parameter
$\dcomm{6}=1$. As a result, the angular dispersion related term $F=0$
and the only two nonzero elements are $C=5.31~\mu$rad/$\mu$m and
$D=9$, resulting in $\xi_{\ind{4}}=266~\mu$rad and $\Delta K =
0.012$~nm$^{-1}$. We note that both the vertical and the horizontal
momentum spreads are smaller than the target specification $\Delta
K=0.02$~nm$^{-1}$.

\subsubsection{Pulse dilation}
\label{dilation}

Bragg diffraction from an asymmetrically cut crystal with angular
dispersion rate $\dirate$ inclines the X-ray intensity front by an
angle $\beta = \arctan(\dirate E)$ resulting in a pulse dilation
$\delta t = \dirate E x/c$ \cite{SL12} along the optical axis
$z$. Here $x$ is the transverse pulse size after the angular
dispersive optics and $c$ is the speed of light in vacuum. This effect
is similar to wavefront inclination by optical diffraction
gratings. The multi-crystal CDDW+W optic has a very large cumulative
angular dispersion rate $\fcomm{6} = 112~\mu$rad/meV (see
Table~\ref{tab2}). The result is an inclination of the pulse intensity
front by $\beta=\arctan(\dirate E)=89.94^{\circ}$ and thus a very
large pulse stretching $\delta t\,=\, \fcomm{6}\, E\, x_{\ind{2}}/\, c
= 190$~ps (equivalent to a 57~mm pulse length). Here
$x_{\ind{2}}=x_{\ind{0}} \mu_{\ind{2}}/\dcomm{6} =56~\mu$m is the
vertical beam size after the CDDW monochromator.

\subsection{Spectrograph}
\label{spectrograph}

Spectral analysis of photons scattered from the sample is another
important component of IXS spectrometers. Unlike monochromators,
spectral analyzers should have a large angular acceptance, capable of
collecting photons from the greatest possible solid angle (limited
only by the required momentum transfer resolution), and with a
spectral resolution matched to that of the monochromator. The spectral
analyzer is usually the most difficult part of IXS spectrometers. In a
standard approach the IXS analyzers measure sequentially one spectral
point after another.  A better strategy is to image the entire or a
large part of the IXS spectra in single shots. Therefore, in the IXS
instrument proposed here, the photon spectra are measured by an X-ray
spectrograph. A spectrograph is an optical instrument that disperses
photons of different energies into distinct directions and space
locations, and images photon spectra on a position-sensitive
detector. Spectrographs consist of collimating, angular dispersive,
and focusing optical elements.  Their principal schematic is shown in
the pictograph of Fig.~\ref{fig001}. Bragg reflecting crystals
arranged in an asymmetric scattering geometry are used as dispersing
elements (DE) of the hard X-ray spectrograph studied here
\cite{Shv11,Shvydko12,SSM13,Shvydko15}.

\begin{figure*}[t!]
\includegraphics[width=0.8\textwidth]{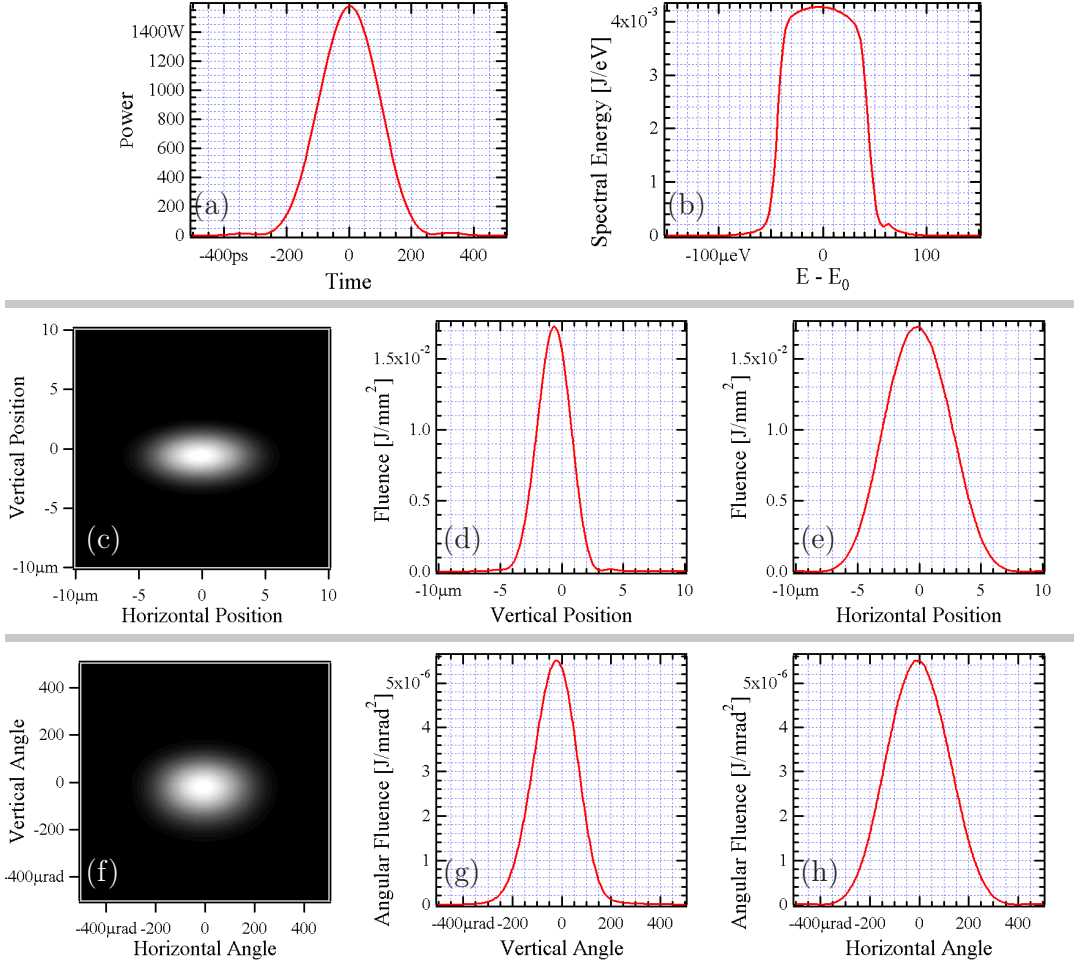}
\caption{Temporal, spectral, spatial, and angular distributions of the
  radiation pulse on the sample ($z=1018$~m in Fig.~\ref{fig001}). (a)
  Pulse power; the pulse duration is $\simeq 225$~ps (FWHM). (b)
  Spectrum; the spectral bandwidth is $\simeq 0.090$~meV (FWHM). (c)
  2D plot of the spatial distribution. (d) Vertical cut through the
  maximum of the fluence distribution; and (e) horizontal cut. The beam
  size on the sample is $3.3~\mu$m~(V)$\times 6.5~\mu$m~(H)
  (FWHM). (f) Angular distribution, 2D plot; (g) vertical cut through
  the maximum of the fluence distribution; and (h) horizontal cut. Beam
  divergence on the sample is $220~\mu$rad~(V)~$\times ~310~\mu$rad~(H)
  (FWHM), corresponding to a $0.01$~nm$^{-1} \times 0.015$~nm$^{-1}$
  transverse momentum spread.}
\label{fig004333}
\end{figure*}

Several optical designs of hard X-ray spectrographs were proposed and
their performances analyzed in Ref.~\cite{Shvydko15}. Spectrographs
with the desired target energy resolution of 0.1~meV and a spectral
window of imaging up to a few tens of meVs were shown to be feasible
for IXS applications.  We refer to Ref.~\cite{Shvydko15} for
details. Here, we only briefly outline a particular spectrograph
design with a DE consisting of three crystals in a CDW arrangement,
schematically shown in Figure~\ref{fig008}(a). Figure~\ref{fig008}(b)
shows the spectrograph's spectral transmission function with a 5.8~meV
wide window of imaging. The sharp line in the same figure represents
the 0.1~meV design resolution.

\begin{table*}
\centering
\begin{tabular}{|l|l|l|l|l|l|l|l|l|l|}
  \hline   \hline
location & $\Delta t$ & $\Delta E$ & $\Delta x $ &  $\Delta x^{\prime}$ & $\Delta K_{\ind{x}}$ &  pulse        &  photons/ & flux  &  spectral    \\
(method)         &        &            & $ \Delta y$ &  $\Delta y^{\prime}$ & $\Delta K_{\ind{y}}$ &  energy       &  pulse    &       &  flux      \\
         & ps     &  meV       & $\mu$m      &   $\mu$rad         & nm$^{-1}$          &  $\mu$J           &  ph/pulse & ph/s  &  ph/s/meV  \\
\hline
\hline
         &        &            &            &            &                   &                   &               &           &              \\
  Undulator exit ($z$=74~m)&   0.014 &    950    &     50     &  1.8         &         0           &     11000      &   $7.5\times 10^{12}$ &  $2.0\times 10^{17}$  &    $2.1\times 10^{14}$ \\
(GENESIS)        &        &            &     50     &  1.8          &        0           &               &           &      &           \\
\hline
        &        &            &            &              &                   &               &           &      &           \\
sample ($z$=1018~m)  &   225   &    0.087   &     3.3    &  220     &       0.01       &  $0.33$    &   $2.3\times 10^{8}$  &   $6.3\times 10^{12}$  &   $7\times 10^{13}$ \\
(SRW wavefront        &        &            &    6.5    &   310       &       0.015            &               &           &      &           \\
 propagation)       &        &            &            &             &                   &               &           &      &           \\
\cline{2-10}
        &        &            &            &             &                   &               &           &      &           \\
sample ($z$=1018~m)  &   190   &    0.09   &     2.4    &   265    &   0.012       &         &                       &                        &                       \\
(ray-transfer        &        &            &     5.5     &   266 &  0.012   &               &           &      &           \\
    matrix)   &        &            &            &             &                   &               &           &      &           \\
  \hline    \hline
\end{tabular}
\caption{Values (FWHM) of X-ray pulse parameters  at different locations along the beamline in HXRSS mode with the UHRIX setup. See text for details. The total transmittance of the optics is 30\% .}
\label{tab4}
\end{table*}

The spectral resolution of the spectrograph is given by
\begin{equation}
\Delta E_{\indrm{S}} =  \frac{\Delta s}{f_{\indrm{C}}} \frac{| \dcomm{n} |}{\fcomm{n} },
\label{eq003}
\end{equation}
derived using the ray-transfer matrix formalism (see Section~\ref{ray}
and Ref.~\cite{Shvydko15}). A large cumulative dispersion rate
$\fcomm{n}$ of the dispersing element, a small cumulative asymmetry
factor $|\dcomm{n}|$, a large focal distance $f_{\indrm{C}}$ of the
collimating optics, and a small source size $\Delta s$ (beam size on
the sample) are advantageous for better spectral resolution. For the
three-crystal CDW dispersing element, with the optical scheme depicted
in Fig.~\ref{fig008}(a), we have $n=3$, $\fcomm{3}=25~\mu$rad/meV, and
$|\dcomm{3}|=0.5$. The target resolution of $\Delta E_{\indrm{S}}
\lesssim 0.1$~meV is attained with $f_{\indrm{C}}=1$~m and $\Delta
s\lesssim 5~\mu$m. The latter is in fact the origin of the target
specification for the focal spot size on the sample discussed in the
beginning of Section~\ref{xrayoptics}. The estimated design value
$x_{\ind{4}}=2.4~\mu$m, see Section \ref{focalspotsize}, is half the
specification value and hence should yield a two times better spectral
resolution than the 0.1~meV at target. For spectral imaging, focusing
onto the detector is required only in one dimension. Hence, with a 2D
position sensitive detector it is possible to simultaneously image the
spectrum of X-rays along the vertical and the momentum transfer
distribution along the horizontal axis.

\subsection{Wavefront propagation through UHRIX optics}
\label{wave}

In this section the design parameters of the UHRIX are verified by
wavefront propagation calculations. Physical optics simulations of the
interaction of X-rays with the various optical elements of
Figure~\ref{fig001} have been performed with the aid of two
programs. The first, GENESIS \cite{GENESIS}, calculates the original
wavefront of the SASE radiation at the exit of the output undulator,
with the results presented in Section~\ref{sec:radi}. The second, SRW
\cite{SRW}, calculates the wavefront after propagation from the
undulator through drift spaces and optical components by using Fourier
optics compatible local propagators. All together, including all
lenses, crystals, and drift spaces, the beamline contains more than
100 elements. Simulations of the diffracting crystals with SRW have
only recently become possible by addition of a new module \cite{SCS14}
which also has been applied to the design of the planned IXS beamline
at NSLS-II \cite{SCSC}.

The temporal, spectral, spatial, and angular radiation pulse
distributions and their parameters at the FEL undulator exit, $z=74$~m
in Fig.~\ref{fig001}, are given in Fig.~\ref{fig00999}.  Radiation
parameters (FWHM) such as pulse duration $\Delta t$, spectral width
$\Delta E$, transverse size $\Delta x$, $\Delta y$; angular spread
$\Delta x^{\prime}$, $\Delta y^{\prime}$, and transverse momentum
spread $\Delta K_{\ind{x}}$, $\Delta K_{\ind{y}}$ are provided in
captions to Fig.~\ref{fig00999} and summarized in Table~\ref{tab4}
together with peak and average flux values. The peak values are also a
result of averaging over hundred runs with GENESIS, as discussed in
Section~\ref{sec:radi}. The average flux values are obtained assuming
a pulse repetition rate of 27~kHz.

\begin{figure*}[t!]
\includegraphics[width=1.0\textwidth]{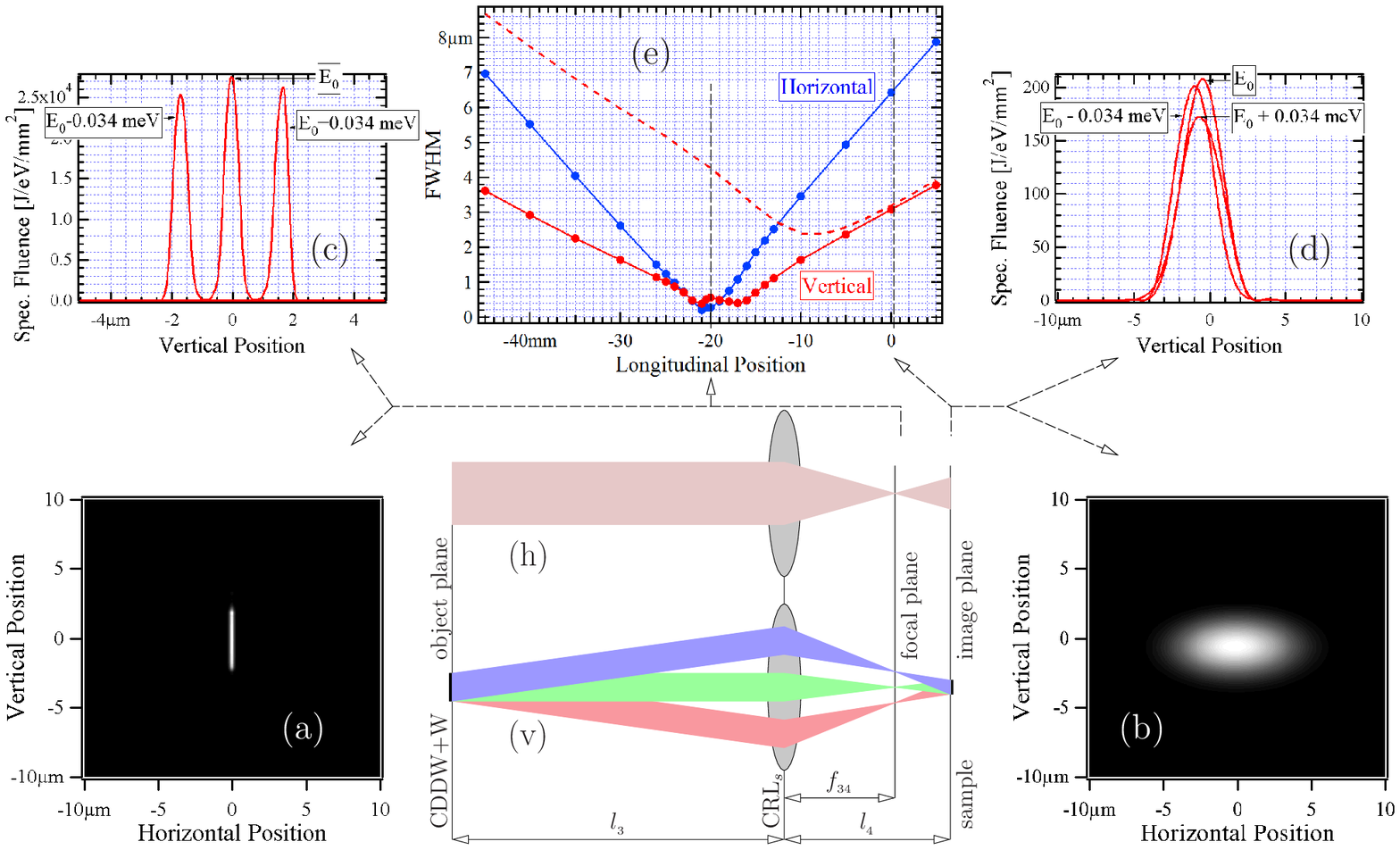}
\caption{Fluence distributions and spot sizes of X-rays at different
  longitudinal positions near the sample.  (a) Fluence distribution
  near the focal plane, and (b) in the sample (image) plane integrated
  over photon energies or pulse duration. (c) Vertical cuts through
  spectral fluence distributions at zero horizontal position for
  different spectral components near the focal plane (-20~mm), and (d)
  in the image plane (0~mm). (e) Vertical and horizontal spot sizes
  (FWHM) for the monochromatic radiation component $E_{\ind{0}}$ as a
  function of longitudinal position along the beam are presented by
  the solid lines. The red dashed curve in (e) represents the vertical
  size integrated over all spectral components. The optical scheme and
  schematic of ray propagation in the \crls\ focusing system are
  presented both in the vertical (v) and horizontal planes (h).  The
  CDDW+W monochromator is in the object plane while the sample is in
  the image plane. }
\label{fig005}
\end{figure*}

Results of the wavefront propagation simulations related to the sample
area are presented graphically in Fig.~\ref{fig004333}.  The temporal,
spectral, spatial, and angular radiation pulse distributions and their
parameters at the sample location (image plane), $z=1018$~m in
Fig.~\ref{fig001}, are provided in captions to Fig.~\ref{fig004333}
and summarized in Table~\ref{tab4} together with the peak and average
flux values on the sample. The calculated radiation parameters at the
sample location are in good agreement with values obtained by the
ray-transfer matrix approach (Section~\ref{ray}) which are shown for
comparison in Table~\ref{tab4}. They are also in agreement with the
target specifications for the UHRIX instrument defined in
Section~\ref{xrayoptics}.

\subsubsection{Spectral, spatial, and angular distribution}

To avoid enlargement of the beam size on the sample due to the angular
dispersion in the CDDW+W monochromator, it was proposed to place this
monochromator in the object plane of the \crls, see
Section~\ref{focalspotsize}. This works perfectly in the geometrical
optics approximation if the monochromator and the \crls\ are assumed
to be point-like. (See Section~\ref{focalspotsize}, and also the
schematics in Fig.~\ref{fig005}(v) and (h).) The question is how well
this works with realistic sizes of monochromator crystals and of the
individual lenses in the \crls , and with non-zero distances between
all these elements. To address these issues, wavefront propagation
simulations have been performed under realistic conditions. Detailed
results are presented in Fig.~\ref{fig005}, showing fluence
distributions and spot sizes of X-rays at different longitudinal
positions near the sample. There are striking differences in the
transverse shape and sizes, integrated over all spectral components,
in the image plane (Fig.~\ref{fig005}(b)) and in the focal plane
(Fig.~\ref{fig005}(a)). There are equally striking differences in the
positions and widths of the vertical beam profiles for different
spectral components in the image plane (Fig.~\ref{fig005}(d)) and in
the focal plane (Fig.~\ref{fig005}(c)).

The widths of the vertical pulse profiles (FWHM) for the monochromatic
component $E_{\ind{0}}$ at different locations are presented in
Fig.~\ref{fig005}(e) by the red solid line. The blue solid line shows
the widths of the horizontal profiles. The smallest widths $\lesssim
0.5~\mu$m of the vertical and horizontal monochromatic pulse profiles
are achieved at $\simeq 21$~mm upstream of the sample position. This
location coincides with the location of the focal plane, which is at a
distance of
$l_{\ind{4}}-f_{\ind{34}}=l_{\ind{4}}^2/(l_{\ind{3}}+l_{\ind{4}})=21$~mm
from the \crls\ center, see sketch in Figs.~\ref{fig005}(v) and
(h). In the image plane the vertical width of approximately $3~\mu$m
is much larger but all monochromatic profiles are almost at the same
position so they probe the same scattering volume, as shown in
Fig.~\ref{fig005}(d). This is in agreement with the ray-transfer
matrix calculations predicting zero linear dispersion in the image
plane, as desired.  In contrast, in the focal plane different
monochromatic components are focused to much smaller sizes ($\simeq
0.5$~$\mu$m) but without spatial overlap, as shown in
Fig.~\ref{fig005}(c).

Figure~\ref{fig005}(v) illustrates the origin of this behavior: Each
monochromatic radiation component emanates from the CDDW+W
monochromator (located in the first approximation in the object plane)
with a very small angular spread $\lesssim 2~\mu$rad. Therefore, with
a virtual source position practically at infinity they are focused
onto the focal plane. Different monochromatic components emanate at
different angles because of strong angular dispersion in the CDDW+W
monochromator that eventually results in a linear dispersion in the
vertical direction of the focal plane but no dispersion in the image
plane, as required for UHRIX. 


The horizontal transverse size of the X-ray pulse is independent of
photon energy, since angular dispersion in the CDDW+W monochromator
takes place only in the vertical plane. The smallest horizontal beam
size is achieved near the focal plane with $\simeq 0.3~\mu$m
\footnote{The small horizontal beam size near the focal plane could be
  used to substantially improve the resolution of the spectrograph,
  see Eq.~\eqref{eq003}. For this, however, its dispersion plane has
  to be oriented horizontally, and the sample placed into the focal
  plane. Alternatively, the dispersion plane of the CDDW+W
  monochromator could be oriented horizontally, to produce a very
  small vertical beamsize on the sample in the focal plane.} (see
Fig.~\ref{fig005}(a) and Fig.~\ref{fig005}(c)). This occurs because of
the very small horizontal angular spread $\lesssim 1~\mu$rad of all
X-ray spectral components emanating from the CDDW+W monochromator.

We note that the best position for the sample is actually neither in
the image plane nor in the focal plane. As follows from the dependence
presented by the dashed line Fig.~\ref{fig005}(e) the smallest
vertical beam size averaged over all spectral components is $\simeq
2.5~\mu$m and it is achived at about -10~mm from the image plane. The
horizontal beam size at the same position is $\simeq 3.5~\mu$m.  We
also note that the extended (realistic 3D model) \crls\ described in
Sec.~\ref{focusingoptics} does not introduce any substantial
differences with respect to the initial simulations with an idealized,
thin \crls.

\subsubsection{Spatiotemporal distributions}

The strong angular dispersion in the CDDW+W monochromator also causes
substantial pulse dilation, as ray-transfer matrix calculations have
shown in Section~\ref{dilation}. Here we present and discuss results
of calculations of the spatiotemporal distributions of the X-ray
pulses obtained by the wavefront propagation simulations.

The pulse duration at the exit of the undulator is only 15~fs (FWHM),
as shown in Fig.~\ref{fig00999}.  The pulse spectral bandwidth is
$\simeq 950$~meV and it is reduced to $\Delta E = 0.09$~meV (FWHM) by
the crystal monochromators. Assuming a Gaussian spectral distribution
after the CDDW+W monochromator, we obtain for the duration of a
Fourier-transform-limited pulse $\Delta t = 4\ln{2}\ \hbar/ \Delta E =
18.2$~ps (FWHM). The results of the calculations shown in
Fig.~\ref{fig004333} predict, however, a more than an order of
magnitude larger pulse duration of $\simeq 225$~ps. This number agrees
well with the duration calculated in Section~\ref{dilation} as a
result of the wave front inclination caused by angular dispersion in
the CDDW+W monochromator.

\subsubsection{Wavefront propagation summary}

The wavefront propagation simulations confirm the soundness of the
optical design of the UHRIX instrument worked out initially by the
ray-transfer matrix approach and dynamical theory calculations. They
also confirm the feasibility of the target specifications. The
simulations show that the spectral flux from the XFEL undulator can be
transported to the sample through the UHRIX X-ray optics with 30\%
efficiency reaching a remarkably high value of $\simeq 7\times
10^{13}$~ph/s/meV. This number exceeds by more than three orders of
magnitude the spectral flux numbers reported for state-of-the-art IXS
instruments at synchrotron radiation facilities
\cite{Baron15arxiv}. Custom designed crystal and focusing optics
ensure that on the sample $\simeq 6.3\times 10^{12}$~ph/s/meV photons
can be concentrated in a spectral band of $0.09$~meV in a spot of
$3.3({\mathrm V})\times 6.5({\mathrm H})~\mu$m$^2$ size and with a
momentum transfer spread of $\lesssim 0.015$~nm$^{-1}$.

\section{Discussion and Conclusions}
\label{conclusions}

This article explores novel opportunities for ultra-high-resolution
IXS (UHRIX) at high repetition rate XFELs unlocked by the recent
demonstration of a conceptually new spectrometer \cite{SSS14} with
unprecedented specifications (0.6 meV spectral resolution and 0.25
nm$^{-1}$ momentum transfer), operating around $9$ keV. Its
exploitation, together with the broadband ultra-high-resolution
imaging spectrograph proposed in \cite{Shvydko15} will make it
possible to fill the energy-momentum gap between high and low
frequency inelastic probes and to provide exciting new opportunities
for studies of dynamics in condensed matter. In particular, UHRIX
experiments can be enabled at the European XFEL, where an increase of
more than three orders of magnitude in average spectral flux is
expected compared to what is available today at synchrotrons. The gain
is due to two main factors: firstly, the high repetition rate of the
European XFEL, owing to the superconducting linac accelerator driver,
which allows up to 27000 X-ray pulses per second, and secondly, the
presence of long undulators, allowing the combined implementation of
hard X-ray self-seeding (HXRSS) and post-saturation tapering
techniques. In particular, a double-chicane HXRSS scheme increases the
signal-to-noise ratio and eases the heat-load on the HXRSS crystals to
a tolerable level. This scheme is expected to yield up to TW-level
X-ray pulses. Simulations of pulse propagation up to the sample
position through the UHRIX optics show that an unprecedented average
spectral flux of $7\times 10^{13}$~ph/s/meV is feasible. The power
delivered to the sample can be as high as 350~W/mm$^2$ and radiation
damage can become a limitation but liquid jets and
scanning setups for solid samples can be employed to circumvent
eventual problems, see Ref.~\cite{MID13} and references therein.

\section{Acknowledgments}

We are grateful to Massimo Altarelli for many useful discussions and
support, and to Thomas Tschentscher, Serguei Molodtsov, Harald Sinn,
Stephen Collins, Giulio Monaco, Alexei Sokolov, Kwang-Je Kim, Kawal
Sawhney, Alexey Suvorov and Igor Zagorodnov for useful discussions and
interest in this work.  Work at the APS was supported by the
U.S. Department of Energy, Office of Science, Office of Basic Energy
Sciences, under Contract No. DE-AC02-06CH11357. The development of SRW
code is supported in part by the US DOE Office of Science, Office of
Basic Energy Sciences under SBIR awards DE-SC0006284 and DE-SC0011237.


\begin{thebibliography}{99}

\expandafter\ifx\csname natexlab\endcsname\relax\def\natexlab#1{#1}\fi
\expandafter\ifx\csname bibnamefont\endcsname\relax
  \def\bibnamefont#1{#1}\fi
\expandafter\ifx\csname bibfnamefont\endcsname\relax
  \def\bibfnamefont#1{#1}\fi
\expandafter\ifx\csname citenamefont\endcsname\relax
  \def\citenamefont#1{#1}\fi
\expandafter\ifx\csname url\endcsname\relax
  \def\url#1{\texttt{#1}}\fi
\expandafter\ifx\csname urlprefix\endcsname\relax\def\urlprefix{URL }\fi
\providecommand{\bibinfo}[2]{#2}
\providecommand{\eprint}[2][]{\url{#2}}

\bibitem[{\citenamefont{Burkel et~al.}(1987)\citenamefont{Burkel, Dorner, and
  Peisl}}]{BDP87}
\bibinfo{author}{\bibfnamefont{E.}~\bibnamefont{Burkel}},
  \bibinfo{author}{\bibfnamefont{B.}~\bibnamefont{Dorner}}, \bibnamefont{and}
  \bibinfo{author}{\bibfnamefont{J.}~\bibnamefont{Peisl}},
  \bibinfo{journal}{Europhys. Lett.} \textbf{\bibinfo{volume}{3}},
  \bibinfo{pages}{957} (\bibinfo{year}{1987}).

\bibitem[{\citenamefont{Burkel}(1991)}]{Burkel}
\bibinfo{author}{\bibfnamefont{E.}~\bibnamefont{Burkel}},
  \emph{\bibinfo{title}{Inelastic Scattering of X rays with Very High Energy
  Resolution}}, vol. \bibinfo{volume}{125} of \emph{\bibinfo{series}{Springer
  Tracts in Modern Physics}} (\bibinfo{publisher}{Springer},
  \bibinfo{address}{Berlin}, \bibinfo{year}{1991}).

\bibitem[{\citenamefont{Sette et~al.}(1998)\citenamefont{Sette, Krisch,
  Masciovecchio, Ruocco, and Monaco}}]{Sette98}
\bibinfo{author}{\bibfnamefont{F.}~\bibnamefont{Sette}},
  \bibinfo{author}{\bibfnamefont{M.~H.} \bibnamefont{Krisch}},
  \bibinfo{author}{\bibfnamefont{C.}~\bibnamefont{Masciovecchio}},
  \bibinfo{author}{\bibfnamefont{G.}~\bibnamefont{Ruocco}}, \bibnamefont{and}
  \bibinfo{author}{\bibfnamefont{G.}~\bibnamefont{Monaco}},
  \bibinfo{journal}{Science} \textbf{\bibinfo{volume}{280}},
  \bibinfo{pages}{1550} (\bibinfo{year}{1998}).

\bibitem[{\citenamefont{Burkel}(2000)}]{Burkel2}
\bibinfo{author}{\bibfnamefont{E.}~\bibnamefont{Burkel}},
  \bibinfo{journal}{Rep. Prog. Phys.} \textbf{\bibinfo{volume}{63}},
  \bibinfo{pages}{171} (\bibinfo{year}{2000}).

\bibitem[{\citenamefont{Krisch and Sette}(2007)}]{KS07}
\bibinfo{author}{\bibfnamefont{M.}~\bibnamefont{Krisch}} \bibnamefont{and}
  \bibinfo{author}{\bibfnamefont{F.}~\bibnamefont{Sette}},
  \emph{\bibinfo{title}{Light Scattering in Solids IX}}
  (\bibinfo{publisher}{Springer}, \bibinfo{address}{Berlin},
  \bibinfo{year}{2007}), vol. \bibinfo{volume}{108} of
  \emph{\bibinfo{series}{Topics in Applied Physics}}, chap.
  \bibinfo{chapter}{Inelastic X-ray Scattering from Phonons}, pp.
  \bibinfo{pages}{317--370}.

\bibitem[{\citenamefont{Monaco}(2015)}]{MonacoIXS15}
\bibinfo{author}{\bibfnamefont{G.}~\bibnamefont{Monaco}},
  \emph{\bibinfo{title}{Synchrotron Radiation}} (\bibinfo{publisher}{Springer
  Berlin Heidelberg}, \bibinfo{year}{2015}), chap. \bibinfo{chapter}{The
  High-Frequency Atomic Dynamics of Disordered Systems Studied by
  High-Resolution Inelastic X-ray Scattering}, pp. \bibinfo{pages}{461--482}.

\bibitem[{\citenamefont{Baron}(2015)}]{Baron15arxiv}
\bibinfo{author}{\bibfnamefont{A.~Q.~R.} \bibnamefont{Baron}},
  \bibinfo{journal}{{arXiv}:1504.01098}  (\bibinfo{year}{2015}).

\bibitem[{\citenamefont{Ashcroft and Mermin}(1976)}]{Askcroft}
\bibinfo{author}{\bibfnamefont{N.~W.} \bibnamefont{Ashcroft}} \bibnamefont{and}
  \bibinfo{author}{\bibfnamefont{N.~D.} \bibnamefont{Mermin}},
  \emph{\bibinfo{title}{Solid State Physics}} (\bibinfo{publisher}{Holt,
  Rinehart and Witson}, \bibinfo{address}{New York}, \bibinfo{year}{1976}).

\bibitem[{\citenamefont{Shvyd'ko et~al.}(2014)\citenamefont{Shvyd'ko, Stoupin,
  Shu, Collins, Mundboth, Sutter, and Tolkiehn}}]{SSS14}
\bibinfo{author}{\bibfnamefont{Y.}~\bibnamefont{Shvyd'ko}},
  \bibinfo{author}{\bibfnamefont{S.}~\bibnamefont{Stoupin}},
  \bibinfo{author}{\bibfnamefont{D.}~\bibnamefont{Shu}},
  \bibinfo{author}{\bibfnamefont{S.~P.} \bibnamefont{Collins}},
  \bibinfo{author}{\bibfnamefont{K.}~\bibnamefont{Mundboth}},
  \bibinfo{author}{\bibfnamefont{J.}~\bibnamefont{Sutter}}, \bibnamefont{and}
  \bibinfo{author}{\bibfnamefont{M.}~\bibnamefont{Tolkiehn}},
  \bibinfo{journal}{Nature Communications} \textbf{\bibinfo{volume}{5:4219}}
  (\bibinfo{year}{2014}).

\bibitem[{\citenamefont{Masciovecchio et~al.}(1996)\citenamefont{Masciovecchio,
  Bergmann, Krisch, Ruocco, Sette, and Verbeni}}]{MBKRSV96}
\bibinfo{author}{\bibfnamefont{C.}~\bibnamefont{Masciovecchio}},
  \bibinfo{author}{\bibfnamefont{U.}~\bibnamefont{Bergmann}},
  \bibinfo{author}{\bibfnamefont{M.}~\bibnamefont{Krisch}},
  \bibinfo{author}{\bibfnamefont{G.}~\bibnamefont{Ruocco}},
  \bibinfo{author}{\bibfnamefont{F.}~\bibnamefont{Sette}}, \bibnamefont{and}
  \bibinfo{author}{\bibfnamefont{R.}~\bibnamefont{Verbeni}},
  \bibinfo{journal}{Nucl. Instrum. Methods Phys. Res. B}
  \textbf{\bibinfo{volume}{117}}, \bibinfo{pages}{339} (\bibinfo{year}{1996}).

\bibitem[{\citenamefont{Said et~al.}(2011)\citenamefont{Said, Sinn, and
  Divan}}]{SSD11}
\bibinfo{author}{\bibfnamefont{A.~H.} \bibnamefont{Said}},
  \bibinfo{author}{\bibfnamefont{H.}~\bibnamefont{Sinn}}, \bibnamefont{and}
  \bibinfo{author}{\bibfnamefont{R.}~\bibnamefont{Divan}},
  \bibinfo{journal}{Journal of Synchrotron Radiation}
  \textbf{\bibinfo{volume}{18}}, \bibinfo{pages}{492} (\bibinfo{year}{2011}).

\bibitem[{\citenamefont{Sette et~al.}(1995)\citenamefont{Sette, Ruocco, Krisch,
  Bergmann, Masciovecchio, Mazzacurati, Signorelli, and Verbeni}}]{SRK95}
\bibinfo{author}{\bibfnamefont{F.}~\bibnamefont{Sette}},
  \bibinfo{author}{\bibfnamefont{G.}~\bibnamefont{Ruocco}},
  \bibinfo{author}{\bibfnamefont{M.}~\bibnamefont{Krisch}},
  \bibinfo{author}{\bibfnamefont{U.}~\bibnamefont{Bergmann}},
  \bibinfo{author}{\bibfnamefont{C.}~\bibnamefont{Masciovecchio}},
  \bibinfo{author}{\bibnamefont{Mazzacurati}},
  \bibinfo{author}{\bibfnamefont{G.}~\bibnamefont{Signorelli}},
  \bibnamefont{and} \bibinfo{author}{\bibfnamefont{R.}~\bibnamefont{Verbeni}},
  \bibinfo{journal}{Phys. Rev. Lett.} \textbf{\bibinfo{volume}{75}},
  \bibinfo{pages}{850} (\bibinfo{year}{1995}).

\bibitem[{\citenamefont{Baron et~al.}(2001)\citenamefont{Baron, Tanaka, Miwa,
  Ishikawa, Mochizuki, Takeshita, Goto, Matsushita, Kimura, Yamamoto
  et~al.}}]{Baron1}
\bibinfo{author}{\bibfnamefont{A.~Q.~R.} \bibnamefont{Baron}},
  \bibinfo{author}{\bibfnamefont{Y.}~\bibnamefont{Tanaka}},
  \bibinfo{author}{\bibfnamefont{D.}~\bibnamefont{Miwa}},
  \bibinfo{author}{\bibfnamefont{D.}~\bibnamefont{Ishikawa}},
  \bibinfo{author}{\bibfnamefont{T.}~\bibnamefont{Mochizuki}},
  \bibinfo{author}{\bibfnamefont{K.}~\bibnamefont{Takeshita}},
  \bibinfo{author}{\bibfnamefont{S.}~\bibnamefont{Goto}},
  \bibinfo{author}{\bibfnamefont{T.}~\bibnamefont{Matsushita}},
  \bibinfo{author}{\bibfnamefont{H.}~\bibnamefont{Kimura}},
  \bibinfo{author}{\bibfnamefont{F.}~\bibnamefont{Yamamoto}},
  \bibnamefont{et~al.}, \bibinfo{journal}{Nucl. Instrum. Methods Phys. Res. A}
  \textbf{\bibinfo{volume}{467-468}}, \bibinfo{pages}{627}
  (\bibinfo{year}{2001}).

\bibitem[{\citenamefont{Sinn et~al.}(2001)\citenamefont{Sinn, Alp, Alatas,
  Barraza, Bortel, Burkel, Shu, Sturhahn, Sutter, Toellner et~al.}}]{SAB01}
\bibinfo{author}{\bibfnamefont{H.}~\bibnamefont{Sinn}},
  \bibinfo{author}{\bibfnamefont{E.}~\bibnamefont{Alp}},
  \bibinfo{author}{\bibfnamefont{A.}~\bibnamefont{Alatas}},
  \bibinfo{author}{\bibfnamefont{J.}~\bibnamefont{Barraza}},
  \bibinfo{author}{\bibfnamefont{G.}~\bibnamefont{Bortel}},
  \bibinfo{author}{\bibfnamefont{E.}~\bibnamefont{Burkel}},
  \bibinfo{author}{\bibfnamefont{D.}~\bibnamefont{Shu}},
  \bibinfo{author}{\bibfnamefont{W.}~\bibnamefont{Sturhahn}},
  \bibinfo{author}{\bibfnamefont{J.}~\bibnamefont{Sutter}},
  \bibinfo{author}{\bibfnamefont{T.}~\bibnamefont{Toellner}},
  \bibnamefont{et~al.}, \bibinfo{journal}{Nucl. Instrum. Methods Phys. Res. A}
  \textbf{\bibinfo{volume}{467-468}}, \bibinfo{pages}{1545}
  (\bibinfo{year}{2001}).

\bibitem[{\citenamefont{Shvyd'ko et~al.}(2013)\citenamefont{Shvyd'ko, Stoupin,
  Mundboth, and Kim}}]{SSM13}
\bibinfo{author}{\bibfnamefont{Y.}~\bibnamefont{Shvyd'ko}},
  \bibinfo{author}{\bibfnamefont{S.}~\bibnamefont{Stoupin}},
  \bibinfo{author}{\bibfnamefont{K.}~\bibnamefont{Mundboth}}, \bibnamefont{and}
  \bibinfo{author}{\bibfnamefont{J.}~\bibnamefont{Kim}},
  \bibinfo{journal}{Phys. Rev. A} \textbf{\bibinfo{volume}{87}},
  \bibinfo{pages}{043835} (\bibinfo{year}{2013}).

\bibitem[{\citenamefont{Stoupin et~al.}(2013)\citenamefont{Stoupin, Shvyd'ko,
  Shu, Blank, Terentyev, Polyakov, Kuznetsov, Lemesh, Mundboth, Collins
  et~al.}}]{SSS13}
\bibinfo{author}{\bibfnamefont{S.}~\bibnamefont{Stoupin}},
  \bibinfo{author}{\bibfnamefont{Y.~V.} \bibnamefont{Shvyd'ko}},
  \bibinfo{author}{\bibfnamefont{D.}~\bibnamefont{Shu}},
  \bibinfo{author}{\bibfnamefont{V.~D.} \bibnamefont{Blank}},
  \bibinfo{author}{\bibfnamefont{S.~A.} \bibnamefont{Terentyev}},
  \bibinfo{author}{\bibfnamefont{S.~N.} \bibnamefont{Polyakov}},
  \bibinfo{author}{\bibfnamefont{M.~S.} \bibnamefont{Kuznetsov}},
  \bibinfo{author}{\bibfnamefont{I.}~\bibnamefont{Lemesh}},
  \bibinfo{author}{\bibfnamefont{K.}~\bibnamefont{Mundboth}},
  \bibinfo{author}{\bibfnamefont{S.~P.} \bibnamefont{Collins}},
  \bibnamefont{et~al.}, \bibinfo{journal}{Opt. Express}
  \textbf{\bibinfo{volume}{21}}, \bibinfo{pages}{30932} (\bibinfo{year}{2013}).

\bibitem[{\citenamefont{Shvyd'ko}(2015)}]{Shvydko15}
\bibinfo{author}{\bibfnamefont{Y.}~\bibnamefont{Shvyd'ko}},
  \bibinfo{journal}{Phys. Rev. A} \textbf{\bibinfo{volume}{91}},
  \bibinfo{pages}{053817} (\bibinfo{year}{2015}).

\bibitem[{\citenamefont{Trigo et~al.}(2013)\citenamefont{Trigo, Fuchs, Chen,
  Jiang, Cammarata, Fahy, Fritz, Gaffney, Ghimire, Higginbotham
  et~al.}}]{TFC13}
\bibinfo{author}{\bibfnamefont{M.}~\bibnamefont{Trigo}},
  \bibinfo{author}{\bibfnamefont{M.}~\bibnamefont{Fuchs}},
  \bibinfo{author}{\bibfnamefont{J.}~\bibnamefont{Chen}},
  \bibinfo{author}{\bibfnamefont{M.~P.} \bibnamefont{Jiang}},
  \bibinfo{author}{\bibfnamefont{M.}~\bibnamefont{Cammarata}},
  \bibinfo{author}{\bibfnamefont{S.}~\bibnamefont{Fahy}},
  \bibinfo{author}{\bibfnamefont{D.~M.} \bibnamefont{Fritz}},
  \bibinfo{author}{\bibfnamefont{K.}~\bibnamefont{Gaffney}},
  \bibinfo{author}{\bibfnamefont{S.}~\bibnamefont{Ghimire}},
  \bibinfo{author}{\bibfnamefont{A.}~\bibnamefont{Higginbotham}},
  \bibnamefont{et~al.}, \bibinfo{journal}{Nature Physics}
  \textbf{\bibinfo{volume}{9}}, \bibinfo{pages}{790–794}
  (\bibinfo{year}{2013}).

\bibitem[{\citenamefont{Kim et~al.}(2008)\citenamefont{Kim, Shvyd\char39{}ko,
  and Reiche}}]{KSR08}
\bibinfo{author}{\bibfnamefont{K.-J.} \bibnamefont{Kim}},
  \bibinfo{author}{\bibfnamefont{Y.}~\bibnamefont{Shvyd\char39{}ko}},
  \bibnamefont{and} \bibinfo{author}{\bibfnamefont{S.}~\bibnamefont{Reiche}},
  \bibinfo{journal}{Phys. Rev. Lett.} \textbf{\bibinfo{volume}{100}},
  \bibinfo{pages}{244802} (\bibinfo{year}{2008}).

\bibitem[{\citenamefont{Lindberg et~al.}(2011)\citenamefont{Lindberg, Kim,
  Shvyd'ko, and Fawley}}]{LSKF11}
\bibinfo{author}{\bibfnamefont{R.~R.} \bibnamefont{Lindberg}},
  \bibinfo{author}{\bibfnamefont{K.-J.} \bibnamefont{Kim}},
  \bibinfo{author}{\bibfnamefont{Y.}~\bibnamefont{Shvyd'ko}}, \bibnamefont{and}
  \bibinfo{author}{\bibfnamefont{W.~M.} \bibnamefont{Fawley}},
  \bibinfo{journal}{Phys. Rev. ST Accel. Beams} \textbf{\bibinfo{volume}{14}},
  \bibinfo{pages}{010701} (\bibinfo{year}{2011}).

\bibitem[{\citenamefont{Maxwell et~al.}(2015)\citenamefont{Maxwell, Arthur,
  Ding, Fawley, Frisch, Hastings, Huang, Krzywinski, Marcus, Kim
  et~al.}}]{XFELO-LCLSII}
\bibinfo{author}{\bibfnamefont{T.~J.} \bibnamefont{Maxwell}},
  \bibinfo{author}{\bibfnamefont{J.}~\bibnamefont{Arthur}},
  \bibinfo{author}{\bibfnamefont{Y.}~\bibnamefont{Ding}},
  \bibinfo{author}{\bibfnamefont{W.~M.} \bibnamefont{Fawley}},
  \bibinfo{author}{\bibfnamefont{J.}~\bibnamefont{Frisch}},
  \bibinfo{author}{\bibfnamefont{J.}~\bibnamefont{Hastings}},
  \bibinfo{author}{\bibfnamefont{Z.}~\bibnamefont{Huang}},
  \bibinfo{author}{\bibfnamefont{J.}~\bibnamefont{Krzywinski}},
  \bibinfo{author}{\bibfnamefont{G.}~\bibnamefont{Marcus}},
  \bibinfo{author}{\bibfnamefont{K.-J.} \bibnamefont{Kim}},
  \bibnamefont{et~al.}, in \emph{\bibinfo{booktitle}{Proceedings of the 2015
  {International} {Particle} {Accelerator} {Conference}}}
  (\bibinfo{publisher}{SLAC Publication: SLAC-PUB-16286},
  \bibinfo{year}{2015}).

\bibitem[{\citenamefont{Emma et~al.}(2010)\citenamefont{Emma, Akre, Arthur,
  Bionta, Bostedt, Bozek, Brachmann, Bucksbaum, Coffee, Decker et~al.}}]{EAA10}
\bibinfo{author}{\bibfnamefont{P.}~\bibnamefont{Emma}},
  \bibinfo{author}{\bibfnamefont{R.}~\bibnamefont{Akre}},
  \bibinfo{author}{\bibfnamefont{J.}~\bibnamefont{Arthur}},
  \bibinfo{author}{\bibfnamefont{R.}~\bibnamefont{Bionta}},
  \bibinfo{author}{\bibfnamefont{C.}~\bibnamefont{Bostedt}},
  \bibinfo{author}{\bibfnamefont{J.}~\bibnamefont{Bozek}},
  \bibinfo{author}{\bibfnamefont{A.}~\bibnamefont{Brachmann}},
  \bibinfo{author}{\bibfnamefont{P.}~\bibnamefont{Bucksbaum}},
  \bibinfo{author}{\bibfnamefont{R.}~\bibnamefont{Coffee}},
  \bibinfo{author}{\bibfnamefont{F.-J.} \bibnamefont{Decker}},
  \bibnamefont{et~al.}, \bibinfo{journal}{Nature Photonics}
  \textbf{\bibinfo{volume}{4}}, \bibinfo{pages}{641 } (\bibinfo{year}{2010}).

\bibitem[{\citenamefont{Ishikawa et~al.}(2012)\citenamefont{Ishikawa, Aoyagi,
  Asaka, Asano, Azumi, Bizen, Ego, Fukami, Fukui, Furukawa et~al.}}]{SACLA}
\bibinfo{author}{\bibfnamefont{T.}~\bibnamefont{Ishikawa}},
  \bibinfo{author}{\bibfnamefont{H.}~\bibnamefont{Aoyagi}},
  \bibinfo{author}{\bibfnamefont{T.}~\bibnamefont{Asaka}},
  \bibinfo{author}{\bibfnamefont{Y.}~\bibnamefont{Asano}},
  \bibinfo{author}{\bibfnamefont{N.}~\bibnamefont{Azumi}},
  \bibinfo{author}{\bibfnamefont{T.}~\bibnamefont{Bizen}},
  \bibinfo{author}{\bibfnamefont{H.}~\bibnamefont{Ego}},
  \bibinfo{author}{\bibfnamefont{K.}~\bibnamefont{Fukami}},
  \bibinfo{author}{\bibfnamefont{T.}~\bibnamefont{Fukui}},
  \bibinfo{author}{\bibfnamefont{Y.}~\bibnamefont{Furukawa}},
  \bibnamefont{et~al.}, \bibinfo{journal}{Nature Photonics}
  \textbf{\bibinfo{volume}{6}}, \bibinfo{pages}{540–544}
  (\bibinfo{year}{2012}).

\bibitem[{\citenamefont{Altarelli et~al.}(2006)\citenamefont{Altarelli,
  Brinkmann, Chergui, Decking, Dobson, Dusterer, Gr{\"u}bel, Graeff, Graafsma,
  Hajdu et~al.}}]{EXFEL-TDR}
\bibinfo{author}{\bibfnamefont{M.}~\bibnamefont{Altarelli}},
  \bibinfo{author}{\bibfnamefont{R.}~\bibnamefont{Brinkmann}},
  \bibinfo{author}{\bibfnamefont{M.}~\bibnamefont{Chergui}},
  \bibinfo{author}{\bibfnamefont{W.}~\bibnamefont{Decking}},
  \bibinfo{author}{\bibfnamefont{B.}~\bibnamefont{Dobson}},
  \bibinfo{author}{\bibfnamefont{S.}~\bibnamefont{Dusterer}},
  \bibinfo{author}{\bibfnamefont{G.}~\bibnamefont{Gr{\"u}bel}},
  \bibinfo{author}{\bibfnamefont{W.}~\bibnamefont{Graeff}},
  \bibinfo{author}{\bibfnamefont{H.}~\bibnamefont{Graafsma}},
  \bibinfo{author}{\bibfnamefont{J.}~\bibnamefont{Hajdu}},
  \bibnamefont{et~al.}, \emph{\bibinfo{title}{XFEL: The European X-ray
  Free-Electron Laser : Technical design report}} (\bibinfo{publisher}{DESY},
  \bibinfo{address}{Hamburg}, \bibinfo{year}{2006}).

\bibitem[{\citenamefont{Madsen et~al.}(2013)\citenamefont{Madsen, Hallmann,
  Roth, and Ansaldi}}]{MID13}
\bibinfo{author}{\bibfnamefont{A.}~\bibnamefont{Madsen}},
  \bibinfo{author}{\bibfnamefont{J.}~\bibnamefont{Hallmann}},
  \bibinfo{author}{\bibfnamefont{T.}~\bibnamefont{Roth}}, \bibnamefont{and}
  \bibinfo{author}{\bibfnamefont{G.}~\bibnamefont{Ansaldi}},
  \bibinfo{type}{Technical Design Report,} \bibinfo{number}{XFEL.EU
  TR-2013-005}, \bibinfo{institution}{{European} {X-ray} {Free-Electron}
  {Laser} {Facility} {GmbH}}, \bibinfo{address}{Hamburg, Germany}
  (\bibinfo{year}{2013}).

\bibitem[{\citenamefont{Geloni et~al.}(2011{\natexlab{a}})\citenamefont{Geloni,
  Kocharyan, and Saldin}}]{GKS11}
\bibinfo{author}{\bibfnamefont{G.}~\bibnamefont{Geloni}},
  \bibinfo{author}{\bibfnamefont{V.}~\bibnamefont{Kocharyan}},
  \bibnamefont{and} \bibinfo{author}{\bibfnamefont{E.}~\bibnamefont{Saldin}},
  \bibinfo{journal}{Journal of Modern Optics} \textbf{\bibinfo{volume}{58}},
  \bibinfo{pages}{1391} (\bibinfo{year}{2011}{\natexlab{a}}).

\bibitem[{\citenamefont{Amann et~al.}(2012)\citenamefont{Amann, Berg, Blank,
  Decker, Ding, Emma, Feng, Frisch, Fritz, Hastings et~al.}}]{HXRSS12}
\bibinfo{author}{\bibfnamefont{J.}~\bibnamefont{Amann}},
  \bibinfo{author}{\bibfnamefont{W.}~\bibnamefont{Berg}},
  \bibinfo{author}{\bibfnamefont{V.}~\bibnamefont{Blank}},
  \bibinfo{author}{\bibfnamefont{F.-J.} \bibnamefont{Decker}},
  \bibinfo{author}{\bibfnamefont{Y.}~\bibnamefont{Ding}},
  \bibinfo{author}{\bibfnamefont{P.}~\bibnamefont{Emma}},
  \bibinfo{author}{\bibfnamefont{Y.}~\bibnamefont{Feng}},
  \bibinfo{author}{\bibfnamefont{J.}~\bibnamefont{Frisch}},
  \bibinfo{author}{\bibfnamefont{D.}~\bibnamefont{Fritz}},
  \bibinfo{author}{\bibfnamefont{J.}~\bibnamefont{Hastings}},
  \bibnamefont{et~al.}, \bibinfo{journal}{Nature Photonics}
  \textbf{\bibinfo{volume}{6}} (\bibinfo{year}{2012}).

\bibitem[{SXS(2014)}]{SXSEED}
\emph{\bibinfo{title}{{XFELSEED}, `{Design} and construction of {Hard} {X-ray}
  {Self-Seeding} {Setups} for the {European} {XFEL}'}},
  \bibinfo{howpublished}{{Project} approved in the framework of the coordinated
  {German-Russian} call for proposals `{Ioffe}-{R\"ontgen} {Institute}'}
  (\bibinfo{year}{2014}).

\bibitem[{\citenamefont{Sprangle et~al.}(1979)\citenamefont{Sprangle, Tang, and
  Manheimer}}]{STC79}
\bibinfo{author}{\bibfnamefont{P.}~\bibnamefont{Sprangle}},
  \bibinfo{author}{\bibfnamefont{C.-M.} \bibnamefont{Tang}}, \bibnamefont{and}
  \bibinfo{author}{\bibfnamefont{W.~M.} \bibnamefont{Manheimer}},
  \bibinfo{journal}{Phys. Rev. Lett.} \textbf{\bibinfo{volume}{43}},
  \bibinfo{pages}{1932} (\bibinfo{year}{1979}).

\bibitem[{\citenamefont{Kroll et~al.}(1981)\citenamefont{Kroll, Morton, and
  Rosenbluth}}]{KMR81}
\bibinfo{author}{\bibfnamefont{N.}~\bibnamefont{Kroll}},
  \bibinfo{author}{\bibfnamefont{P.}~\bibnamefont{Morton}}, \bibnamefont{and}
  \bibinfo{author}{\bibfnamefont{M.}~\bibnamefont{Rosenbluth}},
  \bibinfo{journal}{IEEE J. Quantum Electronics}
  \textbf{\bibinfo{volume}{QE-17}}, \bibinfo{pages}{1436}
  (\bibinfo{year}{1981}).

\bibitem[{\citenamefont{Orzechowski et~al.}(1986)\citenamefont{Orzechowski,
  Anderson, Clark, Fawley, Paul, Prosnitz, Scharlemann, Yarema, Hopkins,
  Sessler et~al.}}]{OAC86}
\bibinfo{author}{\bibfnamefont{T.~J.} \bibnamefont{Orzechowski}},
  \bibinfo{author}{\bibfnamefont{B.~R.} \bibnamefont{Anderson}},
  \bibinfo{author}{\bibfnamefont{J.~C.} \bibnamefont{Clark}},
  \bibinfo{author}{\bibfnamefont{W.~M.} \bibnamefont{Fawley}},
  \bibinfo{author}{\bibfnamefont{A.~C.} \bibnamefont{Paul}},
  \bibinfo{author}{\bibfnamefont{D.}~\bibnamefont{Prosnitz}},
  \bibinfo{author}{\bibfnamefont{E.~T.} \bibnamefont{Scharlemann}},
  \bibinfo{author}{\bibfnamefont{S.~M.} \bibnamefont{Yarema}},
  \bibinfo{author}{\bibfnamefont{D.~B.} \bibnamefont{Hopkins}},
  \bibinfo{author}{\bibfnamefont{A.~M.} \bibnamefont{Sessler}},
  \bibnamefont{et~al.}, \bibinfo{journal}{Phys. Rev. Lett.}
  \textbf{\bibinfo{volume}{57}}, \bibinfo{pages}{2172} (\bibinfo{year}{1986}).

\bibitem[{\citenamefont{Fawley et~al.}(2002)\citenamefont{Fawley, Huang, Kim,
  and Vinokurov}}]{Fawley2002537}
\bibinfo{author}{\bibfnamefont{W.~M.} \bibnamefont{Fawley}},
  \bibinfo{author}{\bibfnamefont{Z.}~\bibnamefont{Huang}},
  \bibinfo{author}{\bibfnamefont{K.-J.} \bibnamefont{Kim}}, \bibnamefont{and}
  \bibinfo{author}{\bibfnamefont{N.~A.} \bibnamefont{Vinokurov}},
  \bibinfo{journal}{Nucl. Instrum. Methods Phys. Res. A}
  \textbf{\bibinfo{volume}{483}}, \bibinfo{pages}{537 } (\bibinfo{year}{2002}).

\bibitem[{\citenamefont{Wang et~al.}(2009)\citenamefont{Wang, Freund, Harder,
  Miner, Murphy, Qian, Shen, and Yang}}]{WFH09}
\bibinfo{author}{\bibfnamefont{X.~J.} \bibnamefont{Wang}},
  \bibinfo{author}{\bibfnamefont{H.~P.} \bibnamefont{Freund}},
  \bibinfo{author}{\bibfnamefont{D.}~\bibnamefont{Harder}},
  \bibinfo{author}{\bibfnamefont{W.~H.} \bibnamefont{Miner}},
  \bibinfo{author}{\bibfnamefont{J.~B.} \bibnamefont{Murphy}},
  \bibinfo{author}{\bibfnamefont{H.}~\bibnamefont{Qian}},
  \bibinfo{author}{\bibfnamefont{Y.}~\bibnamefont{Shen}}, \bibnamefont{and}
  \bibinfo{author}{\bibfnamefont{X.}~\bibnamefont{Yang}},
  \bibinfo{journal}{Phys. Rev. Lett.} \textbf{\bibinfo{volume}{103}},
  \bibinfo{pages}{154801} (\bibinfo{year}{2009}).

\bibitem[{\citenamefont{Geloni et~al.}(2010)\citenamefont{Geloni, Kocharyan,
  and Saldin}}]{GiKoSa10}
\bibinfo{author}{\bibfnamefont{G.}~\bibnamefont{Geloni}},
  \bibinfo{author}{\bibfnamefont{V.}~\bibnamefont{Kocharyan}},
  \bibnamefont{and} \bibinfo{author}{\bibfnamefont{E.}~\bibnamefont{Saldin}},
  \bibinfo{journal}{{arXiv}:1007.2743}  (\bibinfo{year}{2010}),
  \bibinfo{note}{{DESY} 10-108}.

\bibitem[{\citenamefont{Fawley et~al.}(2011)\citenamefont{Fawley, Frisch,
  Huang, Jiao, Nuhn, Pellegrini, Reiche, and Wu}}]{FFH11}
\bibinfo{author}{\bibfnamefont{W.}~\bibnamefont{Fawley}},
  \bibinfo{author}{\bibfnamefont{J.}~\bibnamefont{Frisch}},
  \bibinfo{author}{\bibfnamefont{Z.}~\bibnamefont{Huang}},
  \bibinfo{author}{\bibfnamefont{Y.}~\bibnamefont{Jiao}},
  \bibinfo{author}{\bibfnamefont{H.-D.} \bibnamefont{Nuhn}},
  \bibinfo{author}{\bibfnamefont{C.}~\bibnamefont{Pellegrini}},
  \bibinfo{author}{\bibfnamefont{S.}~\bibnamefont{Reiche}}, \bibnamefont{and}
  \bibinfo{author}{\bibfnamefont{J.}~\bibnamefont{Wu}}, \bibinfo{type}{Tech.
  Rep.}, \bibinfo{institution}{SLAC National Accelerator Laboratory},
  \bibinfo{address}{Menlo Park, CA 94025, USA} (\bibinfo{year}{2011}),
  \bibinfo{note}{{SLAC}-PUB-14616}.

\bibitem[{\citenamefont{Jiao et~al.}(2012)\citenamefont{Jiao, Wu, Cai, Chao,
  Fawley, Frisch, Huang, Nuhn, Pellegrini, and Reiche}}]{JWC12}
\bibinfo{author}{\bibfnamefont{Y.}~\bibnamefont{Jiao}},
  \bibinfo{author}{\bibfnamefont{J.}~\bibnamefont{Wu}},
  \bibinfo{author}{\bibfnamefont{Y.}~\bibnamefont{Cai}},
  \bibinfo{author}{\bibfnamefont{A.~W.} \bibnamefont{Chao}},
  \bibinfo{author}{\bibfnamefont{W.~M.} \bibnamefont{Fawley}},
  \bibinfo{author}{\bibfnamefont{J.}~\bibnamefont{Frisch}},
  \bibinfo{author}{\bibfnamefont{Z.}~\bibnamefont{Huang}},
  \bibinfo{author}{\bibfnamefont{H.-D.} \bibnamefont{Nuhn}},
  \bibinfo{author}{\bibfnamefont{C.}~\bibnamefont{Pellegrini}},
  \bibnamefont{and} \bibinfo{author}{\bibfnamefont{S.}~\bibnamefont{Reiche}},
  \bibinfo{journal}{Phys. Rev. ST Accel. Beams} \textbf{\bibinfo{volume}{15}},
  \bibinfo{pages}{050704} (\bibinfo{year}{2012}).

\bibitem[{\citenamefont{Yang and Shvyd'ko}(2013)}]{YS13}
\bibinfo{author}{\bibfnamefont{X.}~\bibnamefont{Yang}} \bibnamefont{and}
  \bibinfo{author}{\bibfnamefont{Y.}~\bibnamefont{Shvyd'ko}},
  \bibinfo{journal}{Phys. Rev. ST Accel. Beams} \textbf{\bibinfo{volume}{16}},
  \bibinfo{pages}{120701} (\bibinfo{year}{2013}).

\bibitem[{\citenamefont{Reiche}(1999)}]{GENESIS}
\bibinfo{author}{\bibfnamefont{S.}~\bibnamefont{Reiche}},
  \bibinfo{journal}{Nucl. Instrum. Methods Phys. Res. A}
  \textbf{\bibinfo{volume}{429}}, \bibinfo{pages}{243} (\bibinfo{year}{1999}).

\bibitem[{\citenamefont{Chubar and Elleaume}(1998)}]{SRW}
\bibinfo{author}{\bibfnamefont{O.}~\bibnamefont{Chubar}} \bibnamefont{and}
  \bibinfo{author}{\bibfnamefont{P.}~\bibnamefont{Elleaume}},
  \bibinfo{journal}{EPAC-98 Proceedings} pp. \bibinfo{pages}{1177--1179}
  (\bibinfo{year}{1998}).

\bibitem[{\citenamefont{Inagaki et~al.}(2014)\citenamefont{Inagaki, Tanaka,
  Azumi, Hara, Hasegawa, Inubushi, Kameshima, Kimura, R.~Kinjo, Miura
  et~al.}}]{ITA14}
\bibinfo{author}{\bibfnamefont{T.}~\bibnamefont{Inagaki}},
  \bibinfo{author}{\bibfnamefont{T.}~\bibnamefont{Tanaka}},
  \bibinfo{author}{\bibfnamefont{N.}~\bibnamefont{Azumi}},
  \bibinfo{author}{\bibfnamefont{T.}~\bibnamefont{Hara}},
  \bibinfo{author}{\bibfnamefont{T.}~\bibnamefont{Hasegawa}},
  \bibinfo{author}{\bibfnamefont{Y.}~\bibnamefont{Inubushi}},
  \bibinfo{author}{\bibfnamefont{T.}~\bibnamefont{Kameshima}},
  \bibinfo{author}{\bibfnamefont{H.}~\bibnamefont{Kimura}},
  \bibinfo{author}{\bibfnamefont{H.~M.} \bibnamefont{R.~Kinjo}},
  \bibinfo{author}{\bibfnamefont{A.}~\bibnamefont{Miura}},
  \bibnamefont{et~al.}, in \emph{\bibinfo{booktitle}{Proceedings of {FEL} 2014
  {Conference}}} (\bibinfo{address}{Basel}, \bibinfo{year}{2014}),
  \bibinfo{note}{tUC01}.

\bibitem[{\citenamefont{Geloni et~al.}(2011{\natexlab{b}})\citenamefont{Geloni,
  Kocharyan, and Saldin}}]{TWOC}
\bibinfo{author}{\bibfnamefont{G.}~\bibnamefont{Geloni}},
  \bibinfo{author}{\bibfnamefont{V.}~\bibnamefont{Kocharyan}},
  \bibnamefont{and} \bibinfo{author}{\bibfnamefont{E.}~\bibnamefont{Saldin}},
  \bibinfo{journal}{{arXiv}:1109.5112}  (\bibinfo{year}{2011}{\natexlab{b}}),
  \bibinfo{note}{{DESY} 11-165}.

\bibitem[{\citenamefont{Zagorodnov}(2012)}]{S2ER}
\bibinfo{author}{\bibfnamefont{I.}~\bibnamefont{Zagorodnov}}
  (\bibinfo{year}{2012}), \bibinfo{note}{http://www.desy.de/fel-beam/s2e/}.

\bibitem[{\citenamefont{Sinn}(2012)}]{HSINN}
\bibinfo{author}{\bibfnamefont{H.}~\bibnamefont{Sinn}} (\bibinfo{year}{2012}),
  \bibinfo{note}{private communication}.

\bibitem[{\citenamefont{Lengeler et~al.}(1999)\citenamefont{Lengeler, Schroer,
  T\"ummler, Benner, Richwin, Snigirev, Snigireva, and Drakopoulos}}]{LST99}
\bibinfo{author}{\bibfnamefont{B.}~\bibnamefont{Lengeler}},
  \bibinfo{author}{\bibfnamefont{C.}~\bibnamefont{Schroer}},
  \bibinfo{author}{\bibfnamefont{J.}~\bibnamefont{T\"ummler}},
  \bibinfo{author}{\bibfnamefont{B.}~\bibnamefont{Benner}},
  \bibinfo{author}{\bibfnamefont{M.}~\bibnamefont{Richwin}},
  \bibinfo{author}{\bibfnamefont{A.}~\bibnamefont{Snigirev}},
  \bibinfo{author}{\bibfnamefont{I.}~\bibnamefont{Snigireva}},
  \bibnamefont{and}
  \bibinfo{author}{\bibfnamefont{M.}~\bibnamefont{Drakopoulos}},
  \bibinfo{journal}{J. Synchrotron Radiation} \textbf{\bibinfo{volume}{6}},
  \bibinfo{pages}{1153} (\bibinfo{year}{1999}).

\bibitem[{\citenamefont{Shvyd'ko et~al.}(2006)\citenamefont{Shvyd'ko, Lerche,
  Kuetgens, R{\"u}ter, Alatas, and Zhao}}]{SLK06}
\bibinfo{author}{\bibfnamefont{Y.~V.} \bibnamefont{Shvyd'ko}},
  \bibinfo{author}{\bibfnamefont{M.}~\bibnamefont{Lerche}},
  \bibinfo{author}{\bibfnamefont{U.}~\bibnamefont{Kuetgens}},
  \bibinfo{author}{\bibfnamefont{H.~D.} \bibnamefont{R{\"u}ter}},
  \bibinfo{author}{\bibfnamefont{A.}~\bibnamefont{Alatas}}, \bibnamefont{and}
  \bibinfo{author}{\bibfnamefont{J.}~\bibnamefont{Zhao}},
  \bibinfo{journal}{Phys. Rev. Lett.} \textbf{\bibinfo{volume}{97}},
  \bibinfo{pages}{235502} (\bibinfo{year}{2006}).

\bibitem[{\citenamefont{Shvyd'ko et~al.}(2011)\citenamefont{Shvyd'ko, Stoupin,
  Shu, and Khachatryan}}]{ShSS11}
\bibinfo{author}{\bibfnamefont{Y.}~\bibnamefont{Shvyd'ko}},
  \bibinfo{author}{\bibfnamefont{S.}~\bibnamefont{Stoupin}},
  \bibinfo{author}{\bibfnamefont{D.}~\bibnamefont{Shu}}, \bibnamefont{and}
  \bibinfo{author}{\bibfnamefont{R.}~\bibnamefont{Khachatryan}},
  \bibinfo{journal}{Phys. Rev. A} \textbf{\bibinfo{volume}{84}},
  \bibinfo{pages}{053823} (\bibinfo{year}{2011}).

\bibitem[{\citenamefont{Shvyd'ko}(2004)}]{Shvydko-SB}
\bibinfo{author}{\bibfnamefont{Y.}~\bibnamefont{Shvyd'ko}},
  \emph{\bibinfo{title}{X-ray Optics -- High-Energy-Resolution Applications}},
  vol.~\bibinfo{volume}{98} of \emph{\bibinfo{series}{Optical Sciences}}
  (\bibinfo{publisher}{Springer}, \bibinfo{address}{Berlin Heidelberg
  New~York}, \bibinfo{year}{2004}).

\bibitem[{\citenamefont{Snigirev et~al.}(1996)\citenamefont{Snigirev, Kohn,
  Snigireva, and Lengeler}}]{SKSL}
\bibinfo{author}{\bibfnamefont{A.}~\bibnamefont{Snigirev}},
  \bibinfo{author}{\bibfnamefont{V.}~\bibnamefont{Kohn}},
  \bibinfo{author}{\bibfnamefont{I.}~\bibnamefont{Snigireva}},
  \bibnamefont{and} \bibinfo{author}{\bibfnamefont{B.}~\bibnamefont{Lengeler}},
  \bibinfo{journal}{Nature} \textbf{\bibinfo{volume}{384}}, \bibinfo{pages}{49}
  (\bibinfo{year}{1996}).

\bibitem[{\citenamefont{Shvyd'ko}(2011)}]{Shv11}
\bibinfo{author}{\bibfnamefont{Y.}~\bibnamefont{Shvyd'ko}},
  \bibinfo{journal}{arXiv:1110.6662}  (\bibinfo{year}{2011}).

\bibitem[{\citenamefont{Shvyd'ko}(2012)}]{Shvydko12}
\bibinfo{author}{\bibfnamefont{Y.}~\bibnamefont{Shvyd'ko}},
  \bibinfo{journal}{Proc. SPIE, Advances in X-ray/EUV Optics and Components
  VII} \textbf{\bibinfo{volume}{8502}}, \bibinfo{pages}{85020J}
  (\bibinfo{year}{2012}).

\bibitem[{\citenamefont{Henke et~al.}(1993)\citenamefont{Henke, Gullikson, and
  Davis}}]{HGD93}
\bibinfo{author}{\bibfnamefont{B.~L.} \bibnamefont{Henke}},
  \bibinfo{author}{\bibfnamefont{E.~M.} \bibnamefont{Gullikson}},
  \bibnamefont{and} \bibinfo{author}{\bibfnamefont{J.~C.} \bibnamefont{Davis}},
  \bibinfo{journal}{At. Data Nucl. Data Tables} \textbf{\bibinfo{volume}{54}},
  \bibinfo{pages}{181} (\bibinfo{year}{1993}).

\bibitem[{\citenamefont{Kogelnik and Li}(1966)}]{KL66}
\bibinfo{author}{\bibfnamefont{H.}~\bibnamefont{Kogelnik}} \bibnamefont{and}
  \bibinfo{author}{\bibfnamefont{T.}~\bibnamefont{Li}}, \bibinfo{journal}{Appl.
  Opt.} \textbf{\bibinfo{volume}{5}}, \bibinfo{pages}{1550}
  (\bibinfo{year}{1966}).

\bibitem[{\citenamefont{Matsushita and Kaminaga}(1980)}]{MK80-1}
\bibinfo{author}{\bibfnamefont{T.}~\bibnamefont{Matsushita}} \bibnamefont{and}
  \bibinfo{author}{\bibfnamefont{U.}~\bibnamefont{Kaminaga}},
  \bibinfo{journal}{Journal of Applied Crystallography}
  \textbf{\bibinfo{volume}{13}}, \bibinfo{pages}{472} (\bibinfo{year}{1980}).

\bibitem[{\citenamefont{Siegman}(1986)}]{Siegman}
\bibinfo{author}{\bibfnamefont{A.~E.} \bibnamefont{Siegman}},
  \emph{\bibinfo{title}{Lasers}} (\bibinfo{publisher}{University Science
  Books}, \bibinfo{address}{Sausalito, California}, \bibinfo{year}{1986}).

\bibitem[{\citenamefont{Shvyd'ko and Lindberg}(2012)}]{SL12}
\bibinfo{author}{\bibfnamefont{Y.}~\bibnamefont{Shvyd'ko}} \bibnamefont{and}
  \bibinfo{author}{\bibfnamefont{R.}~\bibnamefont{Lindberg}},
  \bibinfo{journal}{Phys. Rev. ST Accel. Beams} \textbf{\bibinfo{volume}{15}},
  \bibinfo{pages}{100702} (\bibinfo{year}{2012}).

\bibitem[{\citenamefont{Sutter et~al.}(2014)\citenamefont{Sutter, Chubar, and
  Suvorov}}]{SCS14}
\bibinfo{author}{\bibfnamefont{J.~P.} \bibnamefont{Sutter}},
  \bibinfo{author}{\bibfnamefont{O.}~\bibnamefont{Chubar}}, \bibnamefont{and}
  \bibinfo{author}{\bibfnamefont{A.}~\bibnamefont{Suvorov}},
  \bibinfo{journal}{Proc. SPIE} \textbf{\bibinfo{volume}{9209}},
  \bibinfo{pages}{92090L} (\bibinfo{year}{2014}).

\bibitem[{\citenamefont{Suvorov et~al.}(2014)\citenamefont{Suvorov, Cai,
  Sutter, and Chubar}}]{SCSC}
\bibinfo{author}{\bibfnamefont{A.}~\bibnamefont{Suvorov}},
  \bibinfo{author}{\bibfnamefont{Y.~Q.} \bibnamefont{Cai}},
  \bibinfo{author}{\bibfnamefont{J.~P.} \bibnamefont{Sutter}},
  \bibnamefont{and} \bibinfo{author}{\bibfnamefont{O.}~\bibnamefont{Chubar}},
  \bibinfo{journal}{Proc. SPIE} \textbf{\bibinfo{volume}{9209}},
  \bibinfo{pages}{92090H} (\bibinfo{year}{2014}).

\end{thebibliography}
\end{document}